\documentclass[paper]{JHEP3}

\usepackage{epsfig}
\usepackage{amsmath}
\usepackage{amssymb}
\usepackage{amsbsy}
\usepackage{enumerate}
\usepackage{bbm}
\usepackage{url}
\usepackage[hang,sf]{subfigure}
\usepackage{cite}
\newcommand{\tmmathbf}[1]{\ensuremath{{\bf #1}}}
\newcommand{\tmtexttt}[1]{{\ttfamily{#1}}}

\def\beq{\begin{equation}}
\def\beqn{\begin{eqnarray}}
\def\eeq{\end{equation}}
\def\eeqn{\end{eqnarray}}

\def\rg{\right\}} 
\def\lg{\left\{} 
\def\({\left(} 
\def\){\right)}

\newdimen\figwidth
\figwidth=\textwidth
\newcommand\captskip{\vskip -0.7cm}

\newcommand\POWHEG{{\tt POWHEG}}
\newcommand\HERWIG{{\tt HERWIG}}
\newcommand\HWpp{{\tt HERWIG++}}
\newcommand\PYTHIA{{\tt PYTHIA}}
\newcommand\SHERPA{{\tt SHERPA}}

\newcommand\BOX{{\tt POWHEG-BOX}}
\newcommand\MCatNLO{{\tt MC@NLO}}

\newcommand\FKS{Frixione, Kunszt and Signer}

\newcommand\sss{\mathchoice%
{\displaystyle}%
{\scriptstyle}%
{\scriptscriptstyle}%
{\scriptscriptstyle}%
}
\newcommand\nplus{\oplus}
\newcommand\nminus{\ominus}
\newcommand\splus{{\sss \nplus}}
\newcommand\sminus{{\sss \nminus}}
\newcommand\splusminus{{\mathchoice%
{\vplusminus\displaystyle}%
{\vplusminus\scriptstyle}%
{\vplusminus\scriptscriptstyle}%
{\vplusminus\scriptscriptstyle}%
}}

\newdimen\hbigcirc
\newdimen\wbigcirc

\newcommand\vplusminus[1]{%
\settoheight{\hbigcirc}{$#1\bigcirc$}%
\settowidth{\wbigcirc}{$#1\bigcirc$}%
\makebox[\wbigcirc]{%
\makebox[0pt]{\rule[0.4\hbigcirc]{0.5\wbigcirc}{0.05\hbigcirc}}%
\makebox[0pt]{\rule[0.1\hbigcirc]{0.5\wbigcirc}{0.05\hbigcirc}}%
\makebox[0pt]{\rule[0.1\hbigcirc]{0.05\wbigcirc}{0.6\hbigcirc}}%
\makebox[0pt]{$#1\bigcirc$}}%
}

\newcommand\Kplus{K_\splus}
\newcommand\Kminus{K_\sminus}

\newcommand\as{\alpha_{\sss\rm S}}

\newcommand\pt{p_{\sss\rm T}}
\newcommand\ptmin{{\pt^{\min}}}
\newcommand\kt{k_{\sss\rm T}}

\newcommand\matB{{\cal B}}

\newcommand\mydot{\!\cdot\!}

\newcommand\CA{C_{\sss\rm A}}

\newcommand\CF{C_{\sss\rm F}}

\newcount\minutes 
\newcount\scratch 
 
\def\timestamp{
\scratch=\time 
\divide\scratch by 60 
\edef\hours{\the\scratch} 
\multiply\scratch by 60 
\minutes=\time 
\advance\minutes by -\scratch 

\today\ --$\,$\hours:\null 
\ifnum\minutes< 10 0\fi 
\the\minutes
}

\preprint{
  IPPP/10/74\\
  DCPT/10/148}
\title{Single-top $\boldsymbol{Wt}$-channel production matched with
  parton showers using the {\tt\bf POWHEG} method}
\author{Emanuele Re\\
  Institute for Particle Physics Phenomenology, Department of Physics\\
  University of Durham, Durham, DH1 3LE, UK\\
  E-mail: \email{emanuele.re@durham.ac.uk}}
\abstract{We present results for the next-to-leading order calculation
  of single-top $Wt$-channel production interfaced to Shower Monte
  Carlo programs, implemented according to the \POWHEG{} method. A
  comparison with \MCatNLO{} is carried out. Results obtained using
  the \PYTHIA{} shower are also shown and the effect of typical cuts
  is briefly discussed.}
\keywords{QCD, Monte Carlo, NLO Computations, Resummation, Collider Physics
}

\begin{document}
\section{Introduction}
\label{sec:introduction}
Top-quark production is one of the most important processes at hadron
colliders. Within the Standard Model, top quarks can be produced in
pairs (via strong interaction) or individually (via electroweak
processes). Top-antitop pair production is known to have the largest
cross section, and has been studied extensively, both experimentally
and theoretically. In fact, the top quark was discovered in 1995 at
the Tevatron in $t\bar{t}$ production events.  The observation of
single-top electroweak production is instead more difficult and it was
announced by the Tevatron experiments more
recently~\cite{Aaltonen:2009jj,Abazov:2009ii}.  This was due not only
to the fact that single-top cross section is smaller than the
$t\bar{t}$ one, but also to the presence of large backgrounds, namely
$W+\mbox{jet}$ and $t\bar{t}$, that required highly non-trivial
analysis strategies.

Although experimentally challenging, single-top electroweak production
is particularly important since it provides a relatively clean place
to study the electroweak properties of the top quark. For instance, it
allows a direct measurement of the $V_{tb}$ CKM matrix
element~\cite{Alwall:2006bx}, which is important in testing the
unitarity of the CKM matrix. Moreover, electroweak-produced top-quarks
are highly polarized, hence angular correlations of top-quark decay
products are particularly sizeable~\cite{Mahlon:1996pn,Mahlon:1999gz},
providing a good probe of the spin structure of the W-t-b
vertex. Finally, since the top-quark mass is close to the scale of
electroweak symmetry breaking, new physics effects could be discovered
in the the top quark sector, and reactions where only a single-top is
produced are particularly sensitive to some BSM models (see, for
example, refs.~\cite{Tait:2000sh,Cao:2007ea}). Single-top production
(in the $Wt$-channel) is also an important background for some
Higgs-boson search channels, such as $H\to W^+
W^-$~\cite{Dittmar:1996ss}.


The hadroproduction of single-top quarks in proton-proton collisions
is traditionally classified according to the partonic processes
present at LO:
the s-channel processes ($q \bar{q} \to t b$) involve the exchange of
a time-like $W$ boson, the $t$-channel processes ($b q \to t
q^{\prime}$) involve the exchange of a space-like $W$ boson, while
associated $Wt$ production ($b g \to t W^-$) involves the production
of a top quark in association with a $W$ boson. The $t$-channel
process is the main source of single-top quarks, both at the Tevatron
and the LHC. At the Tevatron the $Wt$ contribution is negligible and
the $s$-channel cross section is roughly half of the $t$-channel's
one. At the LHC, instead, the $Wt$-production cross section is a
factor 3 less than the $t$-channel's one, while the $s$-channel is
negligible. Therefore, an accurate description of all the three
production channels is important: in particular, at the LHC, the
$Wt$-channel will play a significant role.

Given the above reasons, it is then desirable to reach a high
precision in the theoretical predictions, both for total rates and for
more exclusive distributions.  Higher order corrections to single-top
hadroproduction have been calculated in
refs.~\cite{Bordes:1994ki,Giele:1995kr,Stelzer:1997ns,
  Harris:2002md,Zhu:2002uj,Campbell:2004ch,Campbell:2005bb,
  Cao:2004ap,Cao:2005pq,Kidonakis:2006bu,Campbell:2009ss,
  Campbell:2009gj}.  Nowadays, one of the ways to go beyond this level
of accuracy
is to merge the fixed order accuracy of a NLO calculation with the
(Next-to)-Leading-Logarithm accuracy (and the flexibility) of a
(parton shower) Monte Carlo event generator. At present there are two
methods to interface NLO calculations with parton showers in a
consistent way: \MCatNLO{}~\cite{Frixione:2002ik} and
\POWHEG{}~\cite{Nason:2004rx,Frixione:2007vw}.\footnote{In literature,
  other proposals
  exist~\cite{Nagy:2005aa,Giele:2007di,Lavesson:2008ah}. The
  \MCatNLO{} and \POWHEG{} methods are the only ones where, currently,
  several full processes at lepton and hadron colliders have been
  implemented.} They have been applied to several processes at
lepton~\cite{LatundeDada:2006gx,LatundeDada:2007jg,LatundeDada:2008bv}
and hadron~\cite{Frixione:2003ei,Frixione:2005vw,Nason:2006hf,
  Frixione:2007nw,Alioli:2008gx,Hamilton:2008pd,Frixione:2008yi,
  Alioli:2008tz,Hamilton:2009za,Papaefstathiou:2009sr,
  LatundeDada:2009rr,Alioli:2009je,Nason:2009ai,Weydert:2009vr,Torrielli:2010aw}
colliders, in conjunction with the \HERWIG{}~\cite{Corcella:2000bw},
\PYTHIA{}~\cite{Sjostrand:2006za} and \HWpp{}~\cite{Bahr:2008pv}
parton shower algorithms.\footnote{In~\cite{Butterworth:2010ym} a
  compared study of the effects of different parton shower algorithms
  has also been performed, in case of Higgs-boson production via gluon
  fusion.}\,\footnote{An implementation of the \POWHEG{} method has
  also appeared recently within the \SHERPA{} event
  generator~\cite{Hoeche:2010pf}.} Despite the theoretical formulation
of the two methods being quite different, reasonable agreement has
been found as well as the reason for the (few) differences encountered
(see for example~\cite{Nason:2010ap} for a recent discussion).

In ref.~\cite{Alioli:2009je} the details of the implementation of
single-top $s$- and $t$-channel production processes in the \POWHEG{}
framework have already been described.  The aim of this paper is to
describe the implementation of the $Wt$-channel production process.
Therefore, results presented here can be seen as a completion of the
work described in the previous reference.

This paper is organized as follows. In sec.~\ref{sec:description} we
describe some of the technicalities of the implementation. We also
summarize how the problem of the interference between the NLO
corrections to this process and the $t\bar{t}$ calculation has been
dealt with. Here we anticipate that we have used the same strategy
adopted by the \MCatNLO{} authors, i.e.~two definitions have been
implemented, such that their difference can be considered a measure of
one of the theoretical uncertainties that affects this process.  In
sec.~\ref{sec:results} results are presented: in particular we compare
\POWHEG{} and \MCatNLO{} results for the two aforementioned
definitions of the NLO corrections. We also show results obtained with
typical cuts and results obtained using the \PYTHIA{} shower. Finally,
in sec.~\ref{sec:conclusions} we give our conclusions.

\section{The \POWHEG{} implementation}
\label{sec:description}

To match single-top $Wt$-channel production NLO corrections with a
parton shower using the \POWHEG{} method, the \BOX{} package has been
used. \BOX{} is a program that automates all the steps described in
ref.~\cite{Frixione:2007vw}, turning a NLO calculation into a
\POWHEG{} simulation. The details of how the program works have been
largely described in ref.~\cite{Alioli:2010xd}. In practice, starting
from some of the typical building blocks of the NLO calculation for
the process at hand, the program produces a set of partonic events
ready to be showered by a shower Monte Carlo program.

In this section we present some of the inputs that we calculated and
then coded in the format needed by the \BOX{} package to work. In
particular, we describe how the LO kinematics and the relevant
differential cross sections up to next-to-leading order in the strong
coupling $\as$ have been obtained. All quark masses have been set to
zero (except, of course, the top-quark mass) and the full
Cabibbo-Kobayashi-Maskawa~(CKM) matrix has been taken into account in
the calculation. However, for sake of illustration, through this
section we set the CKM matrix equal to the identity. Therefore, the
$d$-type quark connected to the top and the $W$ boson will be denoted
as $b$.  Furthermore, in this paper we always refer to top-quark
production: anti-top production is obtained simply by charge
conjugation.

With this convention, the LO partonic process for $Wt$ production is
\begin{equation}
\label{eq:wtch_lo}
b + g \to W^-\! + t\,.
\end{equation}
We denote with $\mathcal{B}_{b g}$ the (summed and averaged) squared
amplitude, divided by the flux factor. The process initiated by the
partons $g b$ is also present. For brevity, throughout the paper
processes that can be obtained from the written ones by simply
exchanging the order of the incoming partons will be omitted from the
formulae.

Before giving the relevant formulae, here we stress again that the NLO
corrections to the $Wt$ production channel are not well defined, due
to interference effects with the $t\bar{t}$ process. This problem is
well known, and several approaches have been
introduced~\cite{Belyaev:1998dn,Tait:1999cf,Campbell:2005bb}. To deal
with this theoretical issue, we used the same strategy first described
in the corresponding \MCatNLO{} publication~\cite{Frixione:2008yi} and
later extensively studied in ref.~\cite{White:2009yt}: two definitions
for the NLO corrections have been considered, and the relative
difference between the two can be interpreted as a measure of the
theoretical uncertainty in the definition of genuine NLO corrections
to the $Wt$-channel process. Since this issue concerns the radiative
part of the QCD NLO corrections, a more accurate discussion will be
given in sec.~\ref{sec:real}.

\subsection{Born kinematics}
\label{sec:kinematics}
Following the notation of ref.~\cite{Frixione:2007vw}, we denote with
$\bar{k}_{\splus}$ and $\bar{k}_{\sminus}$ the incoming parton
momenta, aligned along the plus and minus direction of the $z$ axis,
and by $\bar{k}_1$ and $\bar{k}_2$ the outgoing $W$-boson and
top-quark momenta, respectively. The top-quark and $W$-boson masses
are denoted by $m_t$ and $m_W$. During the \POWHEG{}-algorithm step
where the inclusive NLO cross section is evaluated (the calculation of
the $\bar{B}$ function), their virtualities are kept fixed and equal
to their masses.  If $K_\splus$ and $K_\sminus$ are the momenta of the
incoming hadrons, then we have
\begin{equation}
\bar{k}_\splusminus=\bar{x}_\splusminus K_\splusminus \,,
\end{equation}
where $\bar{x}_\splusminus$ are the momentum fractions, and momentum
conservation reads
\begin{equation}
  \bar{k}_\splus + \bar{k}_\sminus = \bar{k}_1 + \bar{k}_2\,.
\end{equation}
We now introduce the variables
\begin{equation}
  \bar{s} = (\bar{k}_\splus+\bar{k}_\sminus)^2 ,\qquad\quad \bar{Y} =
  \frac{1}{2} 
  \log \frac{(\bar{k}_\splus + \bar{k}_\sminus)^0 + (\bar{k}_\splus +
  \bar{k}_\sminus)^3} {(\bar{k}_\splus + \bar{k}_\sminus)^0 -
  (\bar{k}_\splus + \bar{k}_\sminus)^3}\,,
\end{equation}
and $\bar{\theta}$, the angle between the outgoing top quark and the
$\bar{k}_{\splus}$ momentum, as seen in the partonic
center-of-mass~(CM) frame. We denote with $\bar{\phi}$ the azimuthal
angle of the outgoing top quark in the same reference frame.  Since
the differential cross sections do not depend on the overall azimuthal
orientation of the outgoing partons, we set this angle to zero.  At
the end of the generation of an event, we perform a uniform, random
azimuthal rotation of the whole event, in order to cover the whole
final-state phase space.  The set of variables
$\tmmathbf{\bar{\Phi}}_2 \equiv\lg
\bar{s},\bar{Y},\bar{\theta},\bar{\phi}\rg$ fully parameterizes the
Born kinematics. From them, we can reconstruct the momentum fractions
\begin{equation}
  \label{eq:xpxm_bar}  
  \bar{x}_{\splus}  = \sqrt{\frac{\bar{s}}{S}}\, e^{\bar{Y}}, \qquad\quad
  \bar{x}_{\sminus} = \sqrt{\frac{\bar{s}}{S}}\, e^{- \bar{Y}}\,,
\end{equation}
where $S=(\Kplus + \Kminus)^2$ is the squared CM energy of
the hadronic collider.  The outgoing momenta are first reconstructed
in their longitudinal rest frame, where $\bar{Y}=0$. In this frame, their
energies are
\begin{equation}
  \bar{k}_1^0 |_{\sss \bar{Y}=0}=\frac{\bar{s}+m_W^2-m_t^2}{2\sqrt{\bar{s}}}
  \qquad\mbox{and}\qquad
  \bar{k}_2^0 |_{\sss \bar{Y}=0}=\frac{\bar{s}+m_t^2-m_W^2}{2\sqrt{\bar{s}}}\,.
\end{equation}
The two spatial momenta are opposite and their modulus
$|\vec{k}|_{\sss \bar{Y}=0}$ is found by using the on-shell condition
$m_W^2=(\bar{k}_1^0|_{\sss \bar{Y}=0})^{2} - (|\vec{k}|_{\sss
  \bar{Y}=0})^2$.  We fix the top-quark momentum to form an angle
$\bar{\theta}$ with the $\splus$ direction and to have zero azimuth
(i.e.~it lies in the $x z$ plane and has positive $x$ component). Both
$\bar{k}_1$ and $\bar{k}_2$ are then boosted back in the laboratory
frame, with boost rapidity $\bar{Y}$.  The Born phase space, in terms
of these variables, can be written as
\begin{eqnarray}
\label{eq:Bphsp}
d \tmmathbf{\bar{\Phi}}_2 &=& d\bar{x}_\splus \, d\bar{x}_\sminus
(2\pi)^4 \delta^4\!\(\bar{k}_\splus+ \bar{k}_\sminus -\bar{k}_1 -\bar{k}_2\)
\frac{d^3 \bar{k}_1}{(2\pi)^3 2 \bar{k}_1^0} \,
\frac{d^3 \bar{k}_2}{(2\pi)^3 2 \bar{k}_2^0} \nonumber\\
&=&\frac{1}{S}\frac{\beta}{16\pi}\ d\bar{s}\ d\bar{Y}\ d\cos{\bar{\theta}}\
\frac{d\bar{\phi}}{2\pi}\,,
\end{eqnarray}
where
\begin{equation}
  \beta=\sqrt{1-\rho}\,,\qquad\mbox{with}\qquad \rho=\frac{2\, (m_W^2+m_t^2)}{\bar{s}} - \frac{(m_W^2-m_t^2)^2}{\bar{s}^2}\,.
\end{equation}
The content of eq.~(\ref{eq:Bphsp}) and the procedure described above
to define the parton momenta from the variables set
$\tmmathbf{\bar{\Phi}}_2$ have been coded in the subroutine
\tmtexttt{born\_phsp} of the \BOX{} package~\cite{Alioli:2010xd}.

\subsection{Born and virtual contributions}
\label{sec:bornvirtual}
\begin{figure}[htb]
  \begin{center}
    \epsfig{file=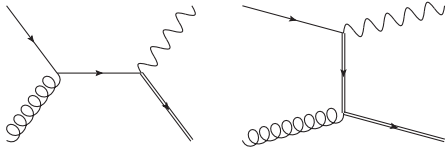,width=0.4\textwidth}
  \end{center}
  \caption{\label{fig:wtch_lo} Feynman diagrams for the LO process $b
    g \to W^- t$. The top quark is denoted with a double line.}
\end{figure}

The squared matrix element for the lowest order subprocess $b g \to
W^- t$ has been obtained with MadGraph~\cite{Alwall:2007st}. It has
been checked with the expression reported in eq.~(3.5) of
ref.~\cite{Frixione:2008yi} and with an independent calculation
performed with FeynCalc~\cite{Mertig:1990an}, starting from the two
Feynman diagrams in fig.~\ref{fig:wtch_lo}. The top-quark width
$\Gamma_t$ has been set here to zero.

To regularize soft and collinear divergences, the \BOX{} package uses
an automated implementation of the subtraction algorithm proposed by
\FKS{}~\cite{Frixione:1995ms,Frixione:1997np}.  The counterterms
needed for the numerical subtraction of the real squared amplitudes
soft divergences are obtained from the color-linked Born squared
amplitudes $\mathcal{B}_{ij}$ defined in eq.~(2.97) of
ref.~\cite{Frixione:2007vw}, using the eikonal approximation.  For the
process at hand, the $\mathcal{B}_{ij}$ matrix is proportional to the
full Born squared amplitude. In fact, since at the lowest order we
have only three color-connected partons, the equality
\begin{equation}
\label{eq:color_cons}
  {\bf T}_\splus + {\bf T}_\sminus +{\bf T}_2 = 0\,
\end{equation}
holds, where ${\bf T}_i$ is the color operator associated with the $i$
parton in the Born process~(\ref{eq:wtch_lo}).  Using the property
\begin{equation}
  {\bf T}_i\mydot{\bf T}_j = 
  \frac{({\bf T}_i + {\bf T}_j)^2 - {\bf T}_i^2 -{\bf T}_j^2}{2}\,,
\end{equation}
and the fact that ${\bf T}_i^2=\CF$ if $i$ is a quark and ${\bf
  T}_i^2=\CA$ if $i$ is a gluon, we obtain ${\matB}_{ij} = (2\CF-\CA)
\matB_{bg}/2$ if neither $i$ or $j$ are gluons, and ${\matB}_{ij} =
\CA \matB_{bg}/2$ if $i$ or $j$ is a gluon.

Collinear counterterms are obtained using the collinear factorization.
However, since in the $Wt$-channel LO process~(\ref{eq:wtch_lo}) an
external gluon is present, the collinear limits associated with this
leg
do not factorize in terms of the Altarelli-Parisi unpolarized
splitting kernels times the Born contribution $\mathcal{B}_{bg}$. In
fact azimuthal correlations in the branching process are present, and
to build a local counterterm the \BOX{} package makes use of the spin
correlated Born cross sections $\mathcal{B}_{\mu\nu}$, defined in
eq.~(2.8) of ref.~\cite{Frixione:2007vw}. The FeynCalc program has
been used to calculate this matrix and to translate the result in a
Fortran routine.

One loop virtual contributions have been calculated and algebraically
reduced to scalar integrals using the Passarino-Veltman algorithm. We
used the same renormalization procedure described in
ref.~\cite{Frixione:2008yi}. We checked that the IR pole structure we
are left with corresponds to the singularities of the real
contributions. The finite part that enters in the soft-virtual
contribution $\mathcal{V}_{bg}$ (eq. (4.6) of
ref.\cite{Alioli:2010xd}) is computed numerically, using the package
QCDloop~\cite{Ellis:2007qk} to evaluate the finite part of the scalar
integrals.

\subsection{Real contributions}
\label{sec:real}
\begin{figure}[htb]
  \begin{center}
    \epsfig{file=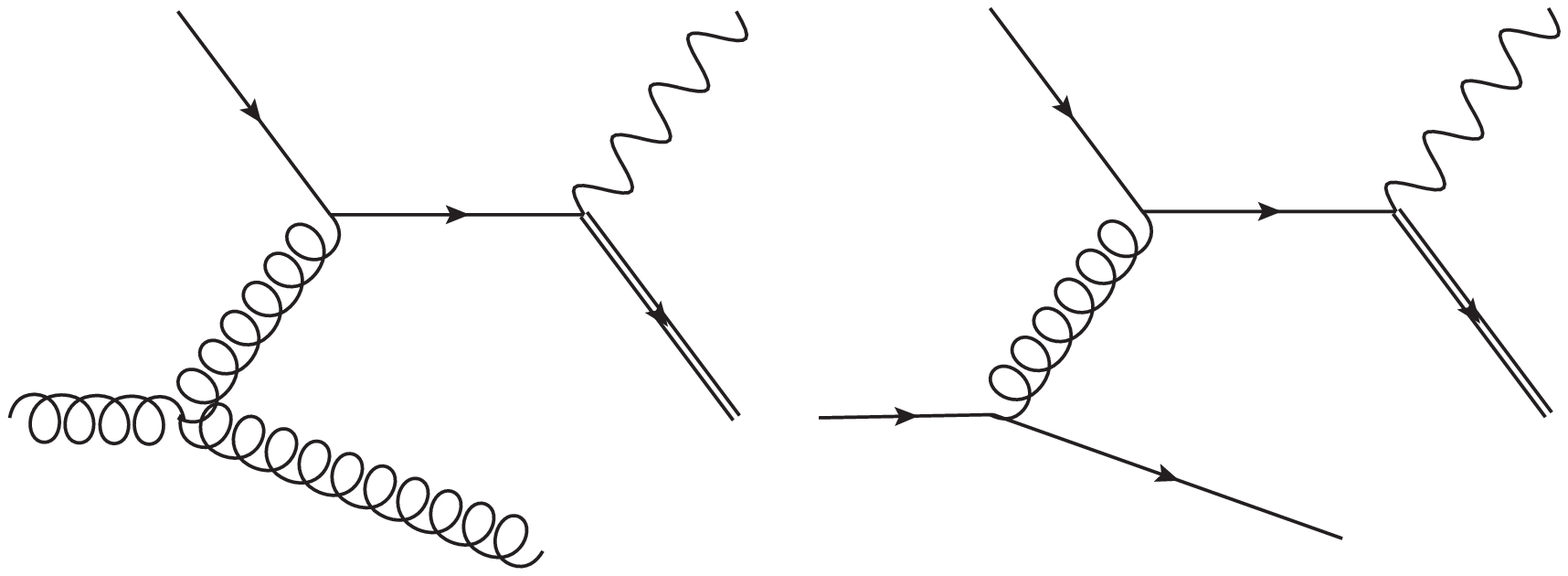,width=0.5\textwidth}\\
    \epsfig{file=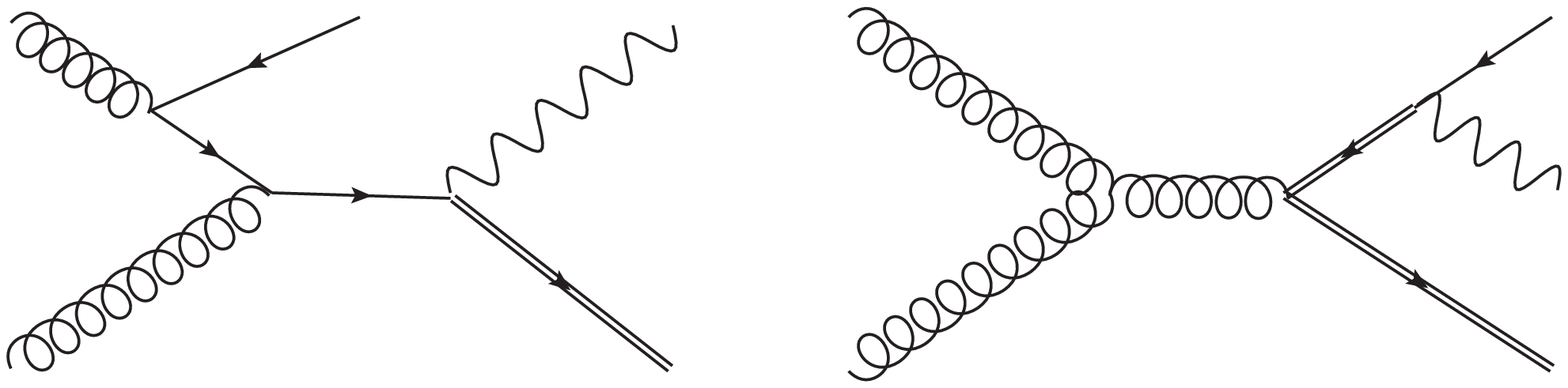,width=0.5\textwidth}\\
    \epsfig{file=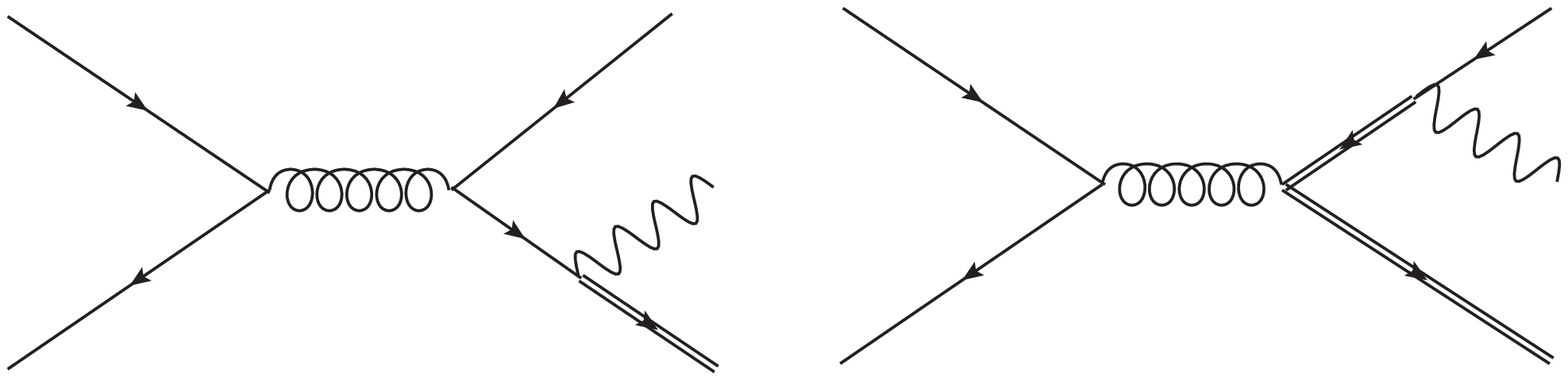,width=0.5\textwidth}
  \end{center}
  \caption{\label{fig:wtch_nlo} Representative Feynman diagrams for
    the processes~(\ref{eq:wtch_r_bg}),~(\ref{eq:wtch_r_bq})
    (up),~(\ref{eq:wtch_r_gg}) (center) and~(\ref{eq:wtch_r_qq})
    (bottom).}
\end{figure}

The real emission corrections can be classified as follows:
\begin{eqnarray}
  \label{eq:wtch_r_bg}
  b + g &\to& W^-\! + t + g\,,\\
  \label{eq:wtch_r_bq}
  b + q(\bar{q}) &\to& W^-\! + t + q(\bar{q})\,\qquad (q\ne b),\\
  \label{eq:wtch_r_gg}
  g + g &\to& W^-\! + t + \bar{b}\,,\\
  \label{eq:wtch_r_qq}
  q + \bar{q} &\to& W^-\! + t + \bar{b}\,.
\end{eqnarray}
We denote the corresponding contributions to the cross section as
$\mathcal{R}_{b g}$, $\mathcal{R}_{b q}$, $\mathcal{R}_{g g}$ and
$\mathcal{R}_{q \bar{q}}$ respectively, where we used again the
standard \POWHEG{} notation first introduced in
ref.~\cite{Frixione:2007vw}. The 3-body phase space is denoted as
$\tmmathbf{\Phi}_3$ and the corresponding momenta as $k_\splus$,
$k_\sminus$, $k_1$, $k_2$ and $k_3$, where $k_3$ is the momentum of
the outgoing light parton (the FKS parton) while the other momenta
correspond to those of the two incoming partons, the $W$-boson and the
top-quark. In fig.~\ref{fig:wtch_nlo} some representative diagrams are
shown.

The processes~(\ref{eq:wtch_r_gg}) and~(\ref{eq:wtch_r_qq}) have a
final state that corresponds to $t\bar{t}$ production followed by the
decay $\bar{t}\to W^- \bar{b}$. A consequence of this fact is the well
known problem of interference between $Wt$ and $t\bar{t}$ production
and it is the reason why QCD NLO corrections to the $Wt$-channel are
not well defined. In the following we will state more precisely the
nature of the problem and explain how we dealt with it.

As one can see from the last four Feynman diagrams in
fig.~\ref{fig:wtch_nlo}, diagrams associated with the
subprocesses~(\ref{eq:wtch_r_gg}) and~(\ref{eq:wtch_r_qq}) can be
divided into two sets. Following the nomenclature of
ref.~\cite{Frixione:2008yi}, we call \emph{doubly-resonant} the
diagrams where a top-quark propagator goes on-shell when the invariant
mass of the system made by the $W$-boson and the outgoing
$\bar{b}$-quark ($m_{W\bar{b}}$) approaches $m_t$. The other diagrams
we call \emph{singly-resonant}. Since these two sets of diagrams have
to be summed at the amplitude level, interference effects are
present. This interference between NLO real corrections to $Wt$ and
lowest order $t\bar{t}$ production (followed by a decay) would not be
a problem if the contamination on $t\bar{t}$-like diagrams was
numerically negligible with respect to the size of singly-resonant
diagrams.  This is certainly not the case when one approaches the
region $(k_W+k_{\,\bar{b}})^2\equiv (k_1+k_3)^2\sim m_t^2$. In fact,
in this region the real corrections to doubly-resonant diagrams become
huge (an internal propagator goes on-shell), and the perturbative
expansion (in power of $\as$) for the NLO corrections to single-top
$Wt$-channel loses its meaning.

To deal with this problem, several approaches are possible:
\begin{itemize}
\item The more drastic approach is to consider that top quarks are not
  detectable particles. This approach would remove any interference
  issue, since processes would be unambiguously classified accordingly
  only to experimentally measurable (QCD) final states: one would have
  the processes $W^+W^-b\bar{b}$ and $W^-W^+b$. At present, the price
  to pay would be to neglect NLO corrections, since these are not
  available for processes where top-quarks are not
  on-shell.\footnote{We also recall that a calculation of NLO
    corrections to the production and the decay of top quarks
    performed in the narrow width approximation (as the one appeared
    in ref.~\cite{Campbell:2005bb}) would not avoid the interference
    problem. To our knowledge, the only calculations that fully
    include top offshellness effects are the ones described in
    ref.~\cite{Pittau:1996rp,Falgari:2010sf} for single-top $s$- and
    $t$-channel production respectively.}
\item An alternative approach is to give a prescription for removing
  the contamination from $t\bar{t}$-like contributions, while keeping
  the top-quark as a final state particle. This task can be
  accomplished using cuts to avoid the doubly-resonant region or
  including an \emph{ad~hoc} subtraction term to remove (or suppress)
  the $t\bar{t}$ contribution.

  The cut strategy was first adopted in ref.~\cite{Belyaev:1998dn},
  where an explicit cut on $m_{W\bar{b}}$ was used. In
  ref.~\cite{Campbell:2005bb} a $b$-jet veto was instead used,
  together with a careful choice of the factorization scale. Moreover,
  in ref.~\cite{Campbell:2005bb} for the first time differential
  results at the NLO were presented and QCD corrections to the
  top-quark decay have been also included, in the narrow-width
  approximation.

  In ref.~\cite{Tait:1999cf} a subtraction term was first introduced
  in the context of an inclusive LO(+LL) calculation. A similar method
  has been adopted in the fully inclusive NLO calculation presented in
  ref.~\cite{Zhu:2002uj}.

  Finally, in refs.~\cite{Frixione:2008yi,White:2009yt} the
  interference problem was reexamined at length, and it was shown that
  a separate treatment of $t\bar{t}$ and $Wt$ production is feasible,
  also when the NLO calculation is matched with a parton shower.
\end{itemize}

In this work we have used the same approach described by the
\MCatNLO{} authors. This relies on the observation that a meaningful
definition of the $Wt$-channel process (as a signal or a background)
is possible only through cuts on final state objects.  If interference
effects with $t\bar{t}$ are negligible \emph{after} these cuts are
applied, then it is possible to consider $Wt$-channel a well defined
process. Since cuts act differently in different phase-space regions,
one needs to quantify the interference between $Wt$ and $t\bar{t}$
locally in the phase-space. To this end, two definitions for the NLO
corrections were given. It was shown that by comparing the two results
it is possible to address the previous question, and it was found
that, for several sets of cuts, the theoretical uncertainty due to
interference effects is typically negligible with respect to other
theoretical errors.  In the following we will discuss briefly the two
subtraction strategies and how they have been implemented within the
\POWHEG{} framework. In sec.~\ref{sec:results} the corresponding
results will be shown.

The two definitions proposed in ref.~\cite{Frixione:2008yi} are known
as \emph{Diagram Removal} (DR) and \emph{Diagram Subtraction} (DS).
Their difference can be better understood by writing a generic
amplitude for the processes~(\ref{eq:wtch_r_gg})
and~(\ref{eq:wtch_r_qq}) as
\begin{equation}
  \label{eq:ampl_wt_tt}
  \mathcal{M}=\mathcal{M}^{Wt} + \mathcal{M}^{t\bar{t}}\,,
\end{equation}
where $\mathcal{M}^{Wt}$ and $\mathcal{M}^{t\bar{t}}$ denote
respectively the sum of all the singly- and doubly-resonant Feynman
diagrams for the partonic subprocess at hand.  In DR one defines the
real contribution $\mathcal{R}$ by eliminating the $t\bar{t}$
contribution $\mathcal{M}^{t\bar{t}}$ from $\mathcal{M}$ before
squaring the amplitude. Instead in DS one keeps the full squared
amplitude but subtracts from it a local counterterm
$\mathcal{C}^{\sss\rm SUB}$ in order to suppress the $t\bar{t}$
contribution at the cross section level.  In this respect, DS can be
seen as a refinement of the method proposed in ref.\cite{Tait:1999cf}.
%
%
Schematically, we have:
\begin{eqnarray}
  \mathcal{R}^{\sss\rm DR} &=& \frac{|\mathcal{M}^{Wt}|^2}{2s}\,,  \label{eq:dr}\\
  \mathcal{R}^{\sss\rm DS} &=& \frac{|\mathcal{M}^{Wt}+\mathcal{M}^{t\bar{t}}|^2 
  - \mathcal{C}^{\sss\rm SUB}}{2s}\,,  \label{eq:ds}
\end{eqnarray}
where $s$ is the squared CM energy.  Some comments are due here:
\begin{itemize}
\item The DR method is not gauge invariant. This issue was studied in
  depth by the authors of ref.\cite{Frixione:2008yi}, and it was shown
  that the impact of gauge dependence in the DR calculation is
  numerically negligible.
\item In the DS approach one wants to build a gauge invariant
  subtraction term that exactly cancels the $t\bar{t}$ contribution
  when the doubly-resonant region is approached. Thus, the subtraction
  term $\mathcal{C}^{\sss\rm SUB}$ has to fulfill the following
  requirements:
  \begin{enumerate}
  \item gauge invariance.
  \item match exactly the $|\mathcal{M}^{t\bar{t}}|^2$ contribution
    when the doubly-resonant region is approached.
  \item possibly fall off (quickly) far from the doubly-resonant
    region.
  \end{enumerate}
  The third requirement is needed to keep the full NLO corrections
  unmodified away from the $t\bar{t}$ peak.  Apart from the three
  requirements above, there is some freedom in the definition of
  $\mathcal{C}^{\sss\rm SUB}$.
\item By taking the difference between eq.~(\ref{eq:ds})
  and~(\ref{eq:dr}), one finds
  \begin{equation}
    \mathcal{R}^{\sss\rm DS}-\mathcal{R}^{\sss\rm DR}=
    \frac{\mathcal{I} + |\mathcal{M}^{t\bar{t}}|^2 - 
      \mathcal{C}^{\sss\rm SUB}}{2 s}\,,
  \end{equation}
  where
  $\mathcal{I}=2\,\Re{(\mathcal{M}^{Wt}\,{\mathcal{M}^{t\bar{t}}}^*)}$.
  Therefore, the difference between results obtained with DR and DS
  can be interpreted as a measure of the size of the interference
  $\mathcal{I}$, provided that the difference
  $|\mathcal{M}^{t\bar{t}}|^2 - \mathcal{C}^{\sss\rm SUB}$ is small.
\end{itemize}

We implemented in \POWHEG{} the two subtraction methods. The squared
amplitudes $|\mathcal{M}|^2$ and $|\mathcal{M}^{Wt}|^2$ have been
obtained using MadGraph. The subtraction term was chosen as in
ref.~\cite{Frixione:2008yi}:
\begin{equation}
  \mathcal{C}^{\sss\rm SUB}(\tmmathbf{\Phi}_3)=\frac{(m_t\Gamma_t)^2}
  {((k_1+k_2)^2-m_t^2)^2 + (m_t\Gamma_t)^2}
  \ {|\mathcal{M}^{t\bar{t}}(\tmmathbf{\Phi}_3^{\prime})|^2}\,,
\end{equation}
where $\tmmathbf{\Phi}_3^{\prime}$ is a point in the 3-body phase
space obtained by reshuffling the $\tmmathbf{\Phi}_3$ kinematics in
order to have $(k_1+k_3)^2=m_t^2$, i.e.~an exactly doubly-resonant
configuration. Hence, the choice of the subtraction term is the same
as the one used by the \MCatNLO{} authors. In fact, in spite of the
aforementioned freedom, as it was already pointed out in
ref.~\cite{Frixione:2008yi}, the choice of the amplitude
$\mathcal{M}^{t\bar{t}}$ evaluated at the point
$\tmmathbf{\Phi}_3^{\prime}$ is unavoidable if one wants to achieve
the exact cancellation of the doubly-resonant contribution while
retaining gauge invariance, since
$\mathcal{M}\sim\mathcal{M}^{t\bar{t}}$ when $(k_1+k_3)^2\to m_t^2$
and gauge invariance is preserved only if the internal $\bar{t}$
propagator is on shell.  The only real freedom is in the choice of the
prefactor, and the Breit-Wigner profile seems the more natural choice
if one wants the difference between $|\mathcal{M}^{t\bar{t}}|^2$ and
$\mathcal{C}^{\sss\rm SUB}$ to be close to zero as much as possible
away from the resonance.

The inclusion of DR in \POWHEG{} is straightforward. The procedure we
adopted for DS is more subtle, since $\mathcal{R}^{\sss\rm DS}$ is not
positive-definite. This affects two steps of the \POWHEG{} method. In
the following we describe how we proceeded in our implementation.  As
usual, we use the standard notation of ref.~\cite{Frixione:2007vw}.
\begin{itemize}
\item Having a real correction that is not always positive-definite
  increases the chances to have regions of $\tmmathbf{\bar{\Phi}}_2$
  where the $\tilde{B}$ function becomes negative. In these cases, the
  radiative event generated with \POWHEG{} starting from the
  underlying-Born configuration $\tmmathbf{\bar{\Phi}}_2$ is
  negative-weighted.  We have checked that this occurrence is rare, so
  that the benefits of having positive-weighted events are not
  spoilt. Moreover, in the \BOX{} package, a procedure (called
  ``folded'' integration) to reduce further the occurrence of these
  events is available, as explained in sec. 4.1 of
  ref.~\cite{Alioli:2010xd}. By using it, we have verified that the
  occurrence of negative-weighted events can be further reduced also
  for the $Wt$-channel case. This also implies that the function
  $\bar{B}$ is positive, as in the other \POWHEG{} implementations.

  Quantitatively, in the event sample we generated to produce the
  plots shown in sec.~\ref{sec:results}, the fraction of
  negative-weighted events was $0.05$. By using the aforementioned
  folded integration, this fraction can be reduced down to $0.005$,
  becoming therefore completely negligible.

\item For 3-body kinematic configurations where $\mathcal{R}^{\sss\rm
    DS}$ is not positive, another problem can occur during the
  generation of the hardest radiation. We recall that, to generate the
  hardest radiation, the \POWHEG{} algorithm works by finding an
  upper-bound for the ratio $\mathcal{R}/\mathcal{B}$, assuming that
  this ratio is always positive. Negative values can therefore spoil
  the accuracy of the method. We have checked that this happens with a
  certain frequency only close to the doubly-resonant region. Since
  this is the kinematic region where the separation of $t\bar{t}$ and
  $Wt$ is already particularly critical,
  we decided to explicitly avoid to generate radiative events when
  close to the doubly-resonant peak.  This has been obtained using a
  theta function that vanishes in this region, i.e.~in DS the
  following substitution is performed when $\mathcal{R}/\mathcal{B}$
  is evaluated:
  \begin{equation}
  \mathcal{R}^{\sss\rm DS} \to \mathcal{R}^{\sss\rm DS} \times \
  \theta\(|m_{W\bar{b}}-m_t| - \kappa \Gamma_t\) \,.
  \end{equation}
  We have tested the effects of this cutoff trying values of order 1
  for $\kappa$, in order to avoid introducing effects from this
  parameter in phase-space regions where it is not needed. The outcome
  of this check is that, at the end of the event generation
  (i.e.~after the shower and the hadronization stage), no problems
  caused by this cutoff are present, since results depend negligibly
  on the value of $\kappa$. The only observable where some dependence
  was observed is the differential distribution of $m_{W\bar{b}}$
  close to the doubly-resonant peak. When values for $\kappa$ in the
  region $3-5$ are used, the difference with the \MCatNLO{} result is
  minimized.  The results shown in sec.~\ref{sec:results} have been
  obtained with $\kappa=3$.
  
  Although we are aware that this solution may be considered
  unappealing, we found it a reasonable choice to handle with the
  problem of having negative values for $\mathcal{R}/\mathcal{B}$ in
  the \POWHEG{} framework. We also recall that the presence of this
  cutoff in our implementation does not affect the issue of negative
  weights, since $\kappa$ is used only in the generation of the
  hardest radiation.
\end{itemize}

We used MadGraph to calculate the term $\mathcal{C}^{\sss\rm SUB}$. We
also notice that to avoid the divergence of the internal top-quark
propagator when $(k_1+k_3)^2\to m_t^2$, in DS a non-vanishing value for
the top width is needed.\footnote{We recall that to obtain the proper
  cancellation of soft and collinear divergences, $\mathcal{M}^{Wt}$
  has to be calculated with $\Gamma_t=0$.}

\section{Results}
\label{sec:results}
In this section we present our results obtained after showering with
\HERWIG{}~6.510 and \PYTHIA{}~6.4.22 the partonic events generated
with \POWHEG{}. We considered top production at the LHC, with an
hadronic center-of-mass energy $\sqrt{S}=10 $ TeV.  All results have
been obtained assuming that the top-quark decays semileptonically
($t\to b\,\bar{\ell}\,\nu$) and that the $W$-boson involved in the
hard scattering decays leptonically ($W^-\to
\ell\,\bar{\nu}$). Branching ratios have been removed, so that plots
are normalized to the total cross section.

We have used the CTEQ6M~\cite{Pumplin:2002vw} set for the parton
distribution functions and the associated value of
\mbox{$\Lambda_{\scriptscriptstyle\overline{\rm
      MS}}^{(5)}=0.226$~GeV}.  Furthermore, as discussed in
refs.~\cite{Frixione:2007vw,Nason:2006hf}, we use a rescaled value
\mbox{$\Lambda_{\scriptscriptstyle\rm{MC}}=
  1.569\,\Lambda_{\scriptscriptstyle\overline{\rm MS}}^{(5)}$} in the
expression for $\as$ appearing in the Sudakov form factors, in order
to achieve next-to-leading logarithmic accuracy.

Although the matrix-element calculation has been performed in the
massless-quark limit, 
the lower cutoff
in the generation of the radiation has been fixed according to the mass of
the emitting quark. The lower bound on the transverse momentum for the
emission off a massless emitter ($u$, $d$, $s$) has been set to the value
$\ptmin = \sqrt{5}\,\Lambda_{\scriptscriptstyle\rm{MC}}$. We instead choose
$\ptmin$ equal to $m_c$ or $m_b$ when the gluon is emitted by a charm or a
bottom quark, respectively.  We set $m_c=1.55$~GeV and $m_b= 4.95$~GeV.

The renormalization and factorization scales have been taken equal to
the transverse momentum of the radiated light parton during the
generation of radiation, in accordance with the \POWHEG{}
formalism. We have also taken into account properly the heavy-flavour
thresholds in the running of $\as$ and in the PDF's, by changing the
number of active flavours when the renormalization or factorization
scales cross a mass threshold.  In the $\bar{B}$ calculation, instead,
$\mu_R$ and $\mu_F$ have been chosen equal to the top-quark mass,
whose value has been fixed to $m_t=175$~GeV. In the DS approach, the
amplitudes where doubly-resonant graphs are present and the
subtraction term $\mathcal{C}^{\sss\rm SUB}$ have been calculated with
$\Gamma_t=1.7$ GeV.

To assess the validity of the approximations and the choices we made,
we compare our results (obtained both with DR and DS) with the
\MCatNLO{} outputs.

\begin{figure}[htb]
  \begin{center}
    \epsfig{file=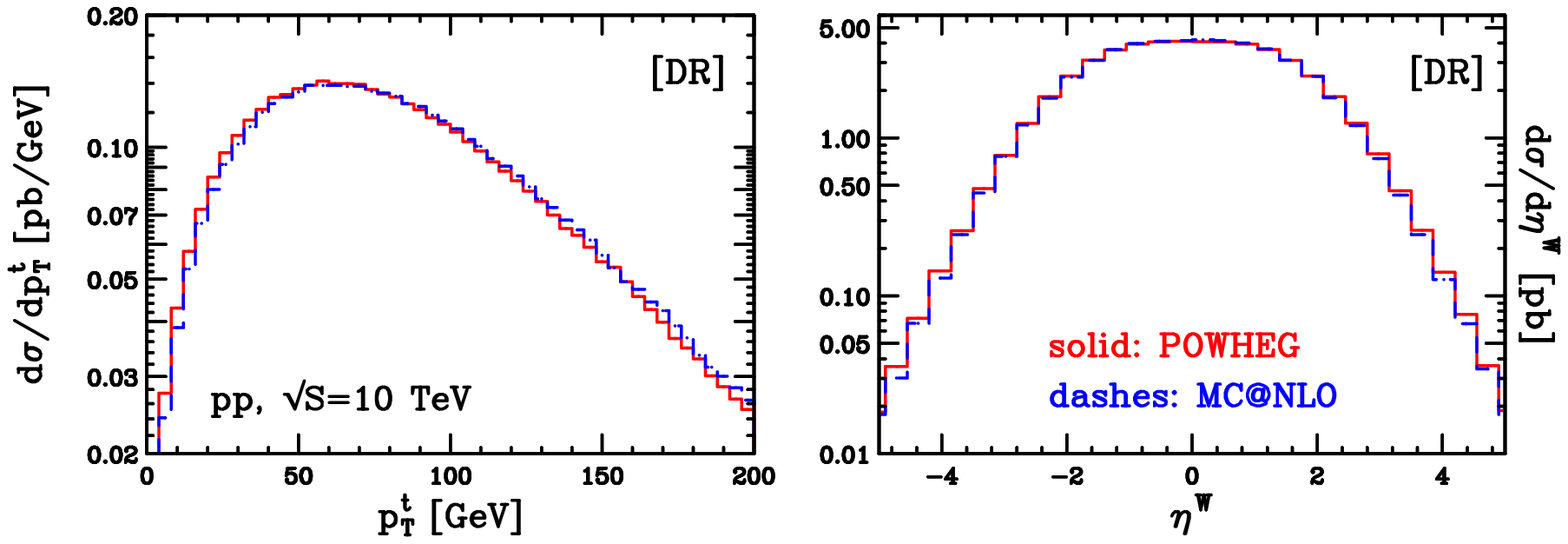,width=\figwidth}\\
    \epsfig{file=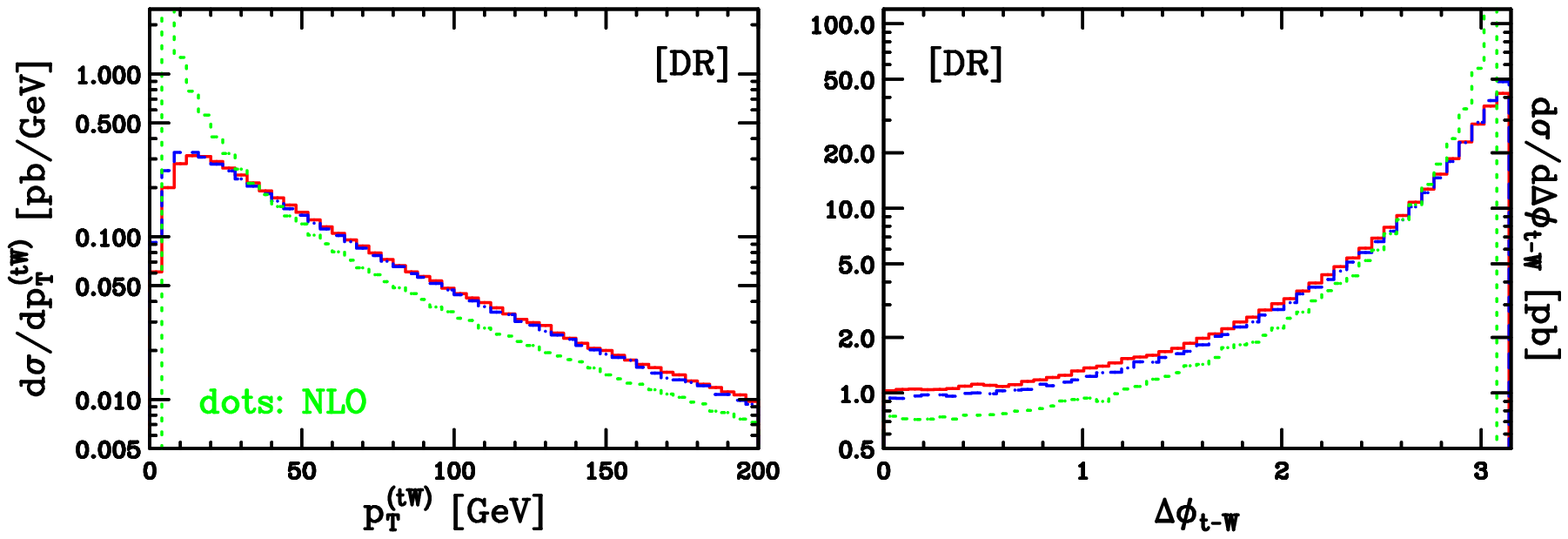,width=\figwidth}
  \end{center}
  \captskip
  \caption{\label{fig:cmp1_DR-POW-MCNLO} Comparisons between \POWHEG{}
    (interfaced to \HERWIG{}) and \MCatNLO{} results at the LHC $pp$
    collider ($\sqrt{S}=10$ TeV), obtained with the DR prescription.
    NLO results are also shown in the lower panel.}
\end{figure}
In fig.~\ref{fig:cmp1_DR-POW-MCNLO} we show a comparison between
\POWHEG{} and \MCatNLO{}, obtained with the DR prescription, without
applying any cut on the final state particles. \POWHEG{} results have
been obtained using the \HERWIG{} parton shower, in order to minimize
differences arising from different shower algorithms and hadronization
models. In the upper panel of fig.~\ref{fig:cmp1_DR-POW-MCNLO} we show
the transverse momentum of the top quark ($\pt^{\,t}$) and the
pseudorapidity of the $W$-boson ($\eta^{\,W}$) produced in the hard
process (i.e.~not the $W$-boson present in the decay chain of the top
quark). As expected, we found very good agreement between the two
results, since the shape of these two distributions is due mainly to
the fixed order result. In the lower panel, we show instead $\pt^{(t
  W\!)}$, the transverse momentum of the system made by the top quark
and the $W$ boson, and $\Delta\phi_{t\mbox{-}W}$, the difference
between the azimuthal angles of the two particles. These two
quantities are significant because their shape is affected by Sudakov
suppression effects due to the resummation performed by parton
showers.  To stress the size of these effects, we also superimposed
the fixed order (NLO) prediction to the last two plots.  We observe
good agreement between showered results, and the expected difference
with the NLO curve, where the cancellation of soft and collinear
divergences takes place only at the edge of the distributions (at
$\pt^{(t W\!)}=0$ and $\Delta\phi_{t\mbox{-}W}=\pi$ respectively).
\begin{figure}[htb]
  \begin{center}
    \epsfig{file=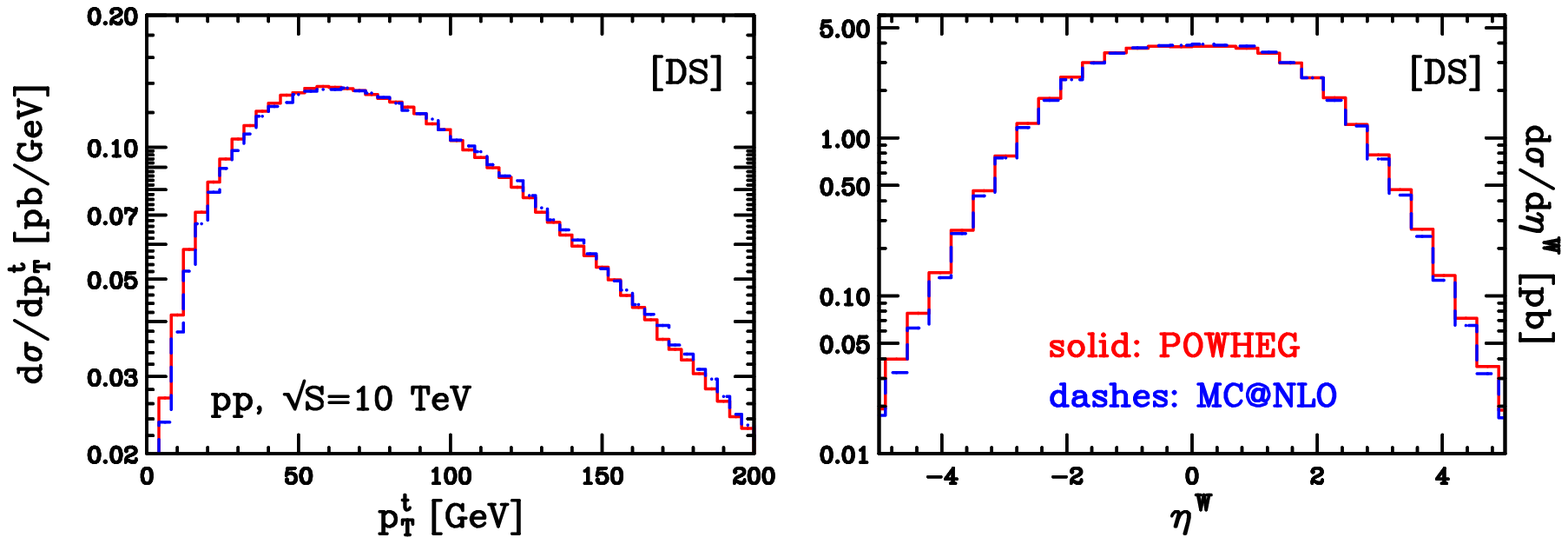,width=\figwidth}\\
    \epsfig{file=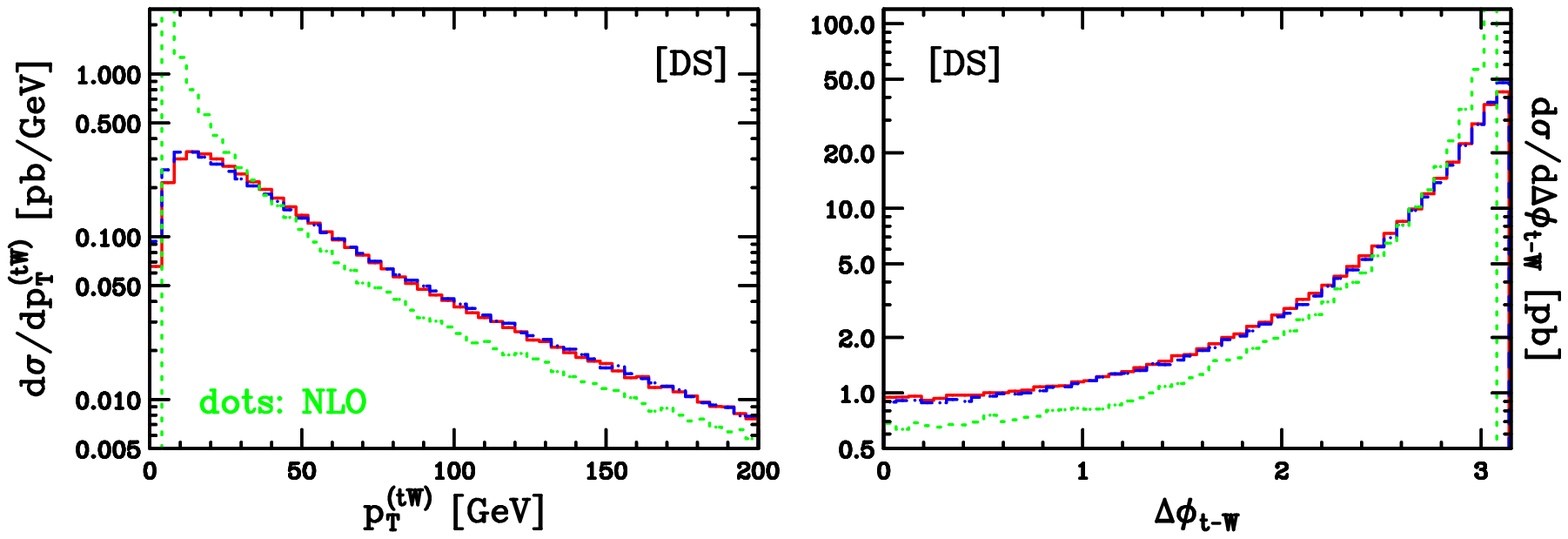,width=\figwidth}
  \end{center}
  \captskip
  \caption{\label{fig:cmp1_DS-POW-MCNLO} Comparisons between \POWHEG{}
    (interfaced to \HERWIG{}) and \MCatNLO{} results at the LHC $pp$
    collider ($\sqrt{S}=10$ TeV), obtained with the DS prescription.
    NLO results are also shown in the lower panel.}
\end{figure}

In fig.~\ref{fig:cmp1_DS-POW-MCNLO} we show the same set of plots,
obtained with the DS prescription. The same considerations made above
are valid also for this case. We recall that the plots shown in
figs.~\ref{fig:cmp1_DR-POW-MCNLO} and~\ref{fig:cmp1_DS-POW-MCNLO} have
been obtained using the top-quark and the $W$-boson momenta extracted
from the parton shower history. Therefore, these quantities are not
measurable in a real detector. Nevertheless, since they are useful to
check the implementation, we have shown the corresponding results.

\begin{figure}[htb]
  \begin{center}
    \epsfig{file=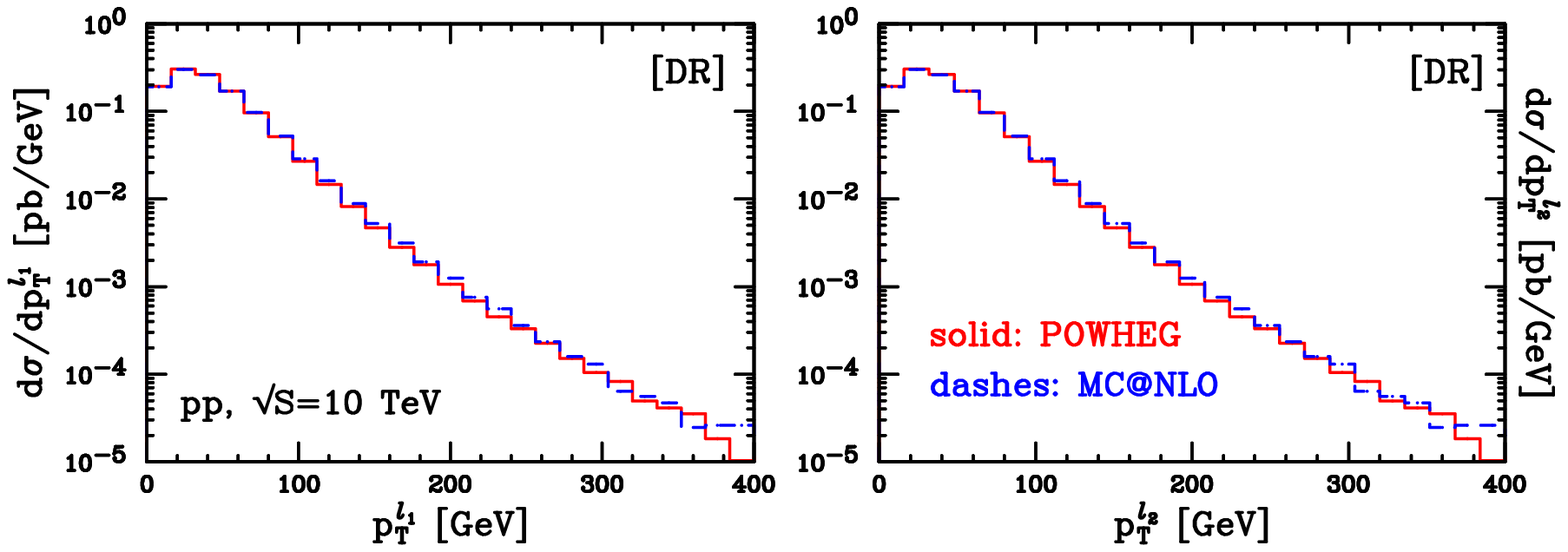,width=\figwidth}\\
    \epsfig{file=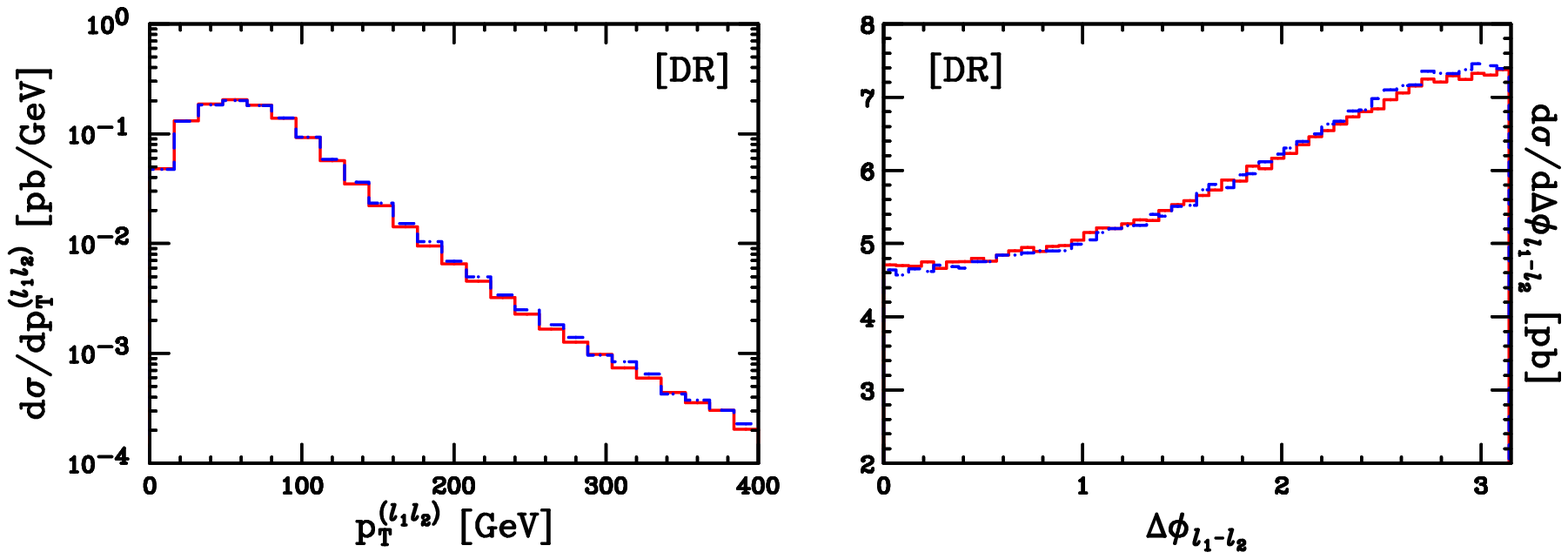,width=\figwidth}
  \end{center}
  \captskip
  \caption{\label{fig:cmp3_DR-POW-MCNLO} Comparisons between \POWHEG{}
    (interfaced to \HERWIG{}) and \MCatNLO{} results at the LHC $pp$
    collider ($\sqrt{S}=10$ TeV), obtained with the DR prescription,
    for leptonic quantities.}
\end{figure}

\begin{figure}[htb]
  \begin{center}
    \epsfig{file=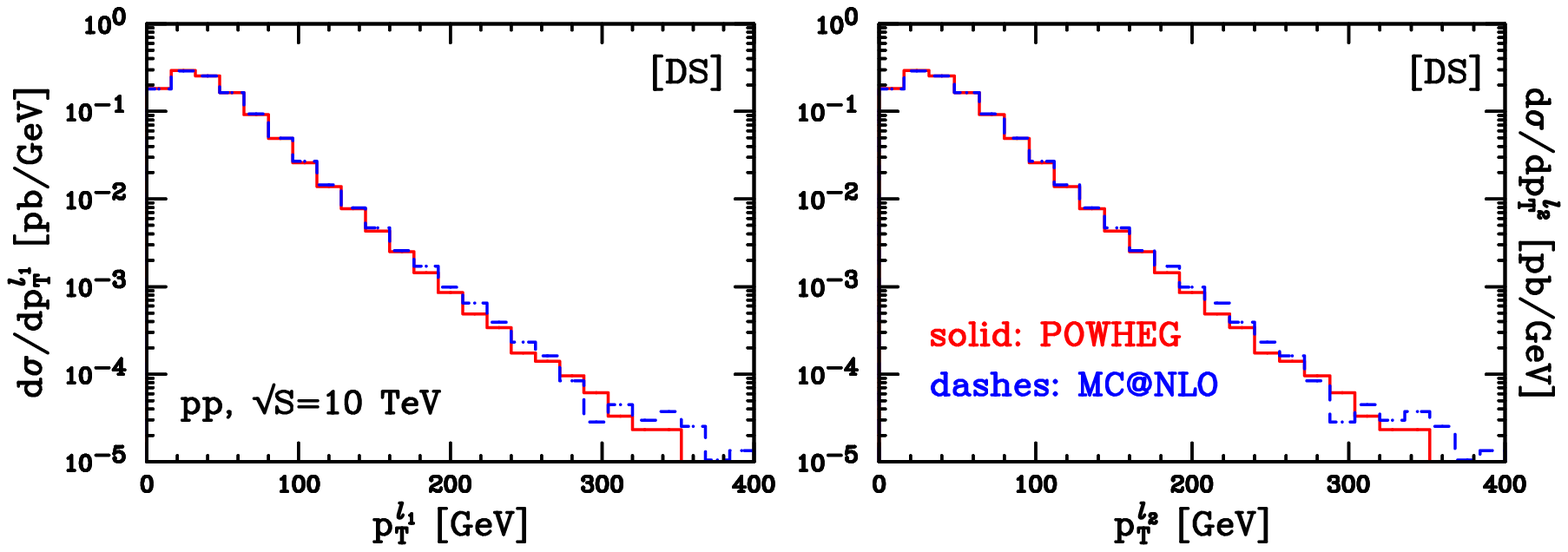,width=\figwidth}\\
    \epsfig{file=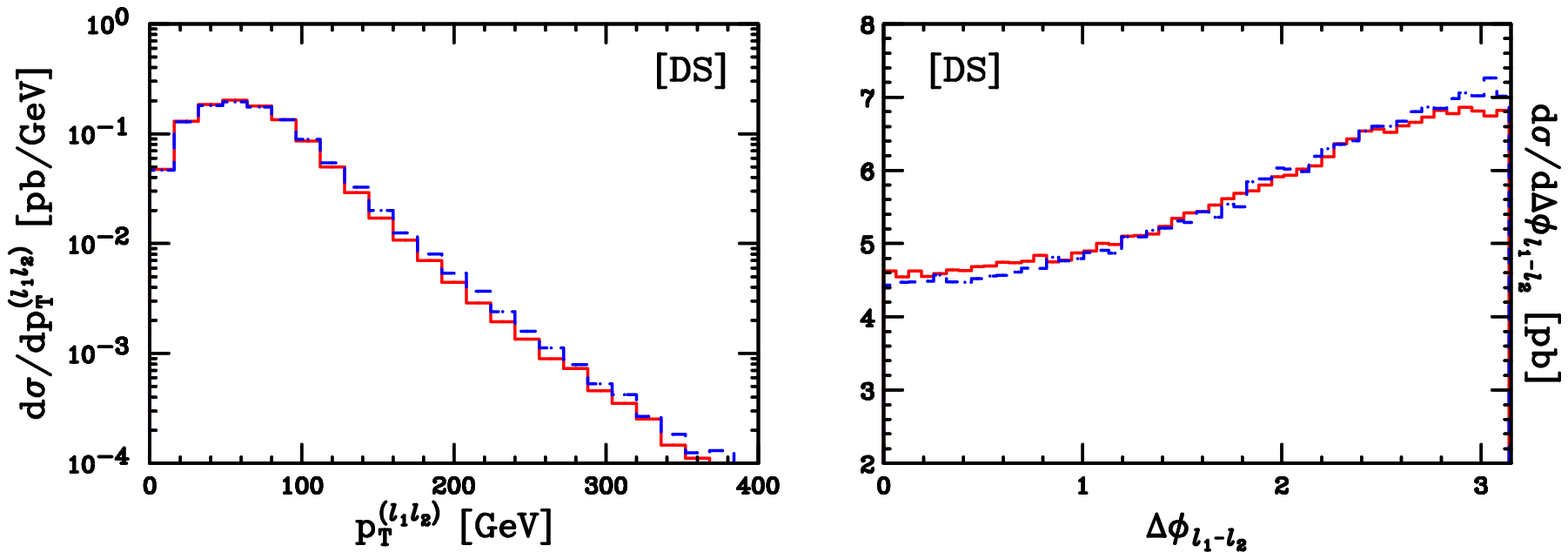,width=\figwidth}
  \end{center}
  \captskip
  \caption{\label{fig:cmp3_DS-POW-MCNLO} Comparisons between \POWHEG{}
    (interfaced to \HERWIG{}) and \MCatNLO{} results at the LHC $pp$
    collider ($\sqrt{S}=10$ TeV), obtained with the DS prescription,
    for leptonic quantities.}
\end{figure}

We have also included the generation of the top-quark and the
$W$-boson decay products, according to the method originally proposed
in ref.~\cite{Frixione:2007zp}. This enables the generation of events
in which spin correlation effects in the production-decay stage are
taken into account with leading-order accuracy.  In
figs.~\ref{fig:cmp3_DR-POW-MCNLO} and~\ref{fig:cmp3_DS-POW-MCNLO} we
show a comparison between \MCatNLO{} and \POWHEG{} for some relevant
leptonic distributions. We plot the transverse momenta of the hardest
($\pt^{\ell_1}$) and the second hardest ($\pt^{\ell_2}$) charged
lepton in the event. We also show $\pt^{(\ell_1 \ell_2\!)}$, the
transverse momentum of the system made by $\ell_1$ and $\ell_2$, and
$\Delta\phi_{\ell_1\mbox{-}\ell_2}$, the difference between the
azimuthal angles of the two leptons. Here, again, we found very good
agreement in both the DR and the DS case.  We recall that a quantity
like $\Delta\phi_{\ell_1\mbox{-}\ell_2}$ is sensitive to
spin-correlation effects, as we have verified by running the code with
the decay-generation procedure switched off and letting the shower
perform isotropic decays.

In ref.~\cite{Campbell:2005bb} the use of a $b$-veto as a method to
discriminate between the $Wt$ and the $t\bar{t}$ processes was
proposed. This idea was also reconsidered in
ref.~\cite{Frixione:2008yi}, where its applicability was studied in
the context of a NLO result merged with a parton shower.  We have
performed a similar exercise using the results obtained with \POWHEG{}
and showered with \HERWIG{}. The $b$-veto condition is defined as
follows: after having sorted in decreasing $\pt$ all the $b$-flavoured
hadrons present in the final state, an event is kept if the
second-hardest $b$-flavoured hadron among those with central
pseudorapidity ($\eta^B\le 2.5$) has $\pt^B<\pt^{(b\rm\mbox{-}veto)}$,
otherwise it is discarded.\footnote{If all the $b$-hadrons in the
  central pseudorapidity region have $\pt^B<\pt^{(b\rm\mbox{-}veto)}$,
  or if the second-hardest $b$-hadron has a large pseudorapidity, the
  event is kept.} In our analysis procedure we chose
$\pt^{(b\rm\mbox{-}veto)}=50$ GeV.

As expected, we observe that the difference between DR and DS results
is reduced when the $b$-veto is applied: in fact, although we have not
performed a detailed study including also uncertainties due to scale
variations, we found that the ratios between the total NLO cross
sections using DR and DS are as follows:
\begin{equation}
  \frac{\sigma_{(\rm\sss DS)}}{\sigma_{(\rm\sss DR)}}= 0.95\,, \ \ 
  \frac{\sigma_{(\rm\sss DS)}^{(b\rm\mbox{-}veto)}}{\sigma_{(\rm\sss DR)}^{(b\rm\mbox{-}veto)}}= 0.98\,.
\end{equation}

\begin{figure}[htb]
  \begin{center}
    \epsfig{file=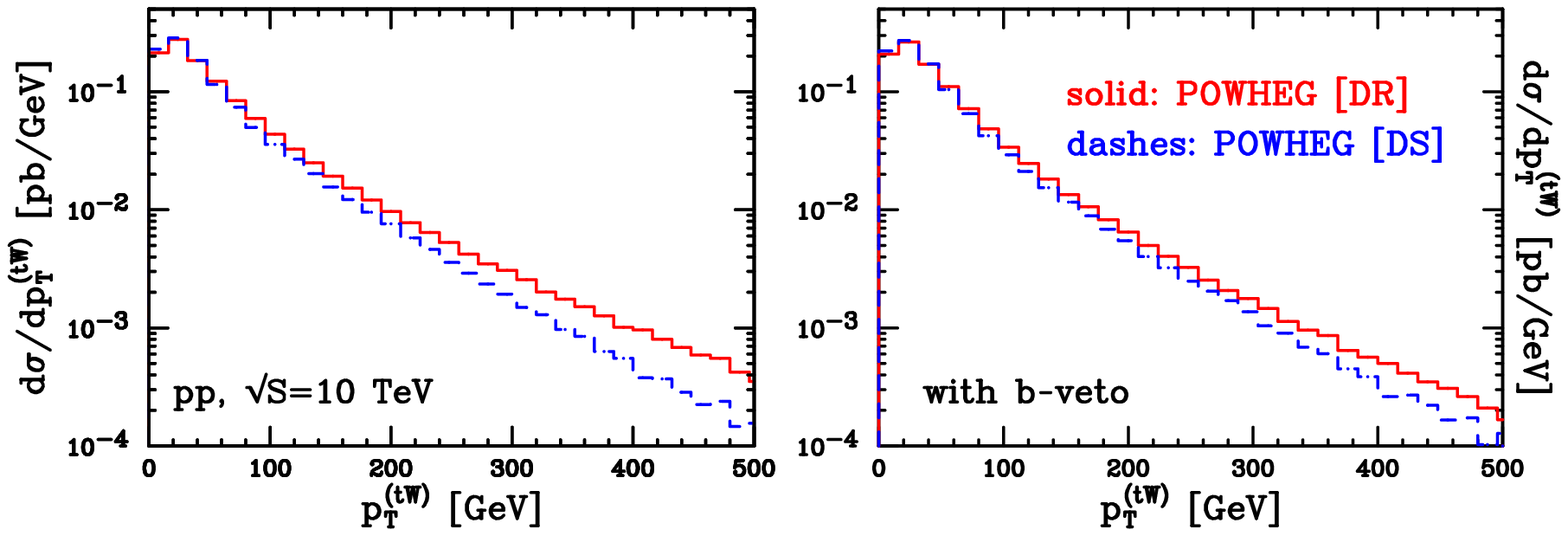,width=\figwidth}\\
    \epsfig{file=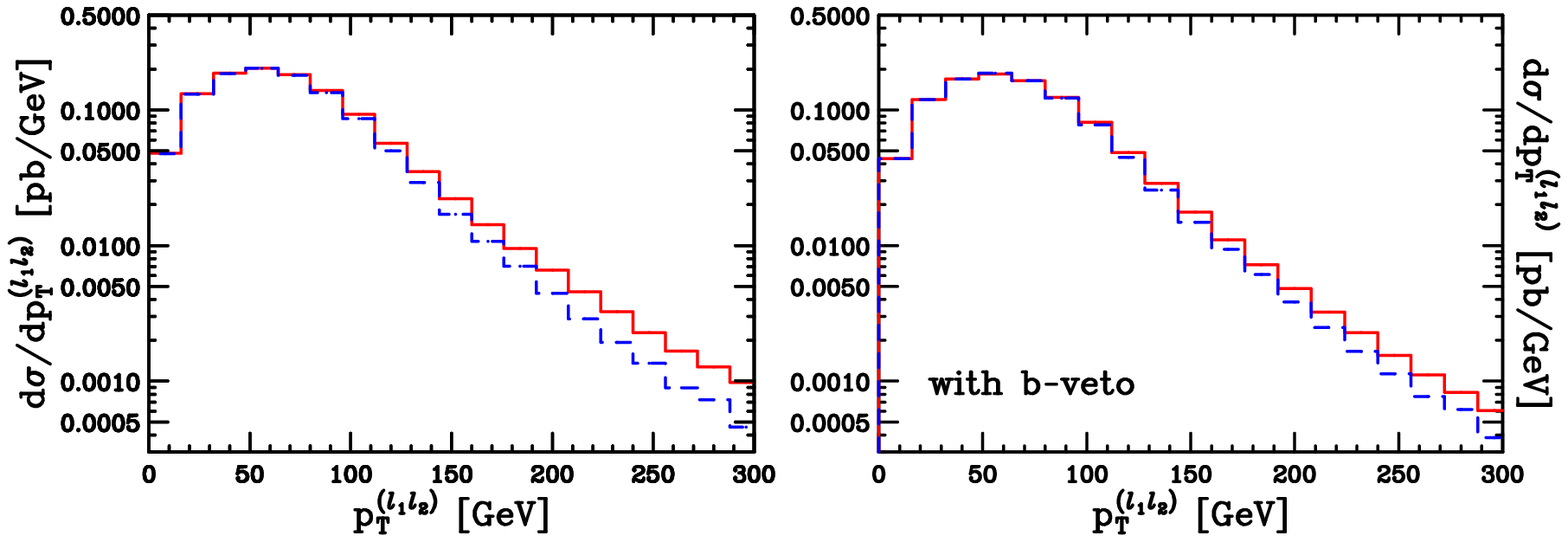,width=\figwidth}
  \end{center}
  \captskip
  \caption{\label{fig:cmpbveto_POWHER-DR-DS} Comparisons between
    \POWHEG{} (interfaced to \HERWIG{}) results obtained with the DR
    and DS prescriptions.  Plots in the right panel have been obtained
    with the $b$-veto described in the text.}
\end{figure}

As the \MCatNLO{} authors already pointed out, despite the fact that
the $b$-veto reduces the difference between DR and DS total cross
sections, it is not guaranteed that all the differential distributions
are affected in the same way. To address this question, in
fig.~\ref{fig:cmpbveto_POWHER-DR-DS} we show the effect of the
$b$-veto cut for two transverse-momentum spectra. In the upper panel
we plot the results for $\pt^{(t W\!)}$, obtained with the DR and the
DS procedures, before and after imposing the $b$-veto. The curves in
the upper-left panel are obtained without cuts, those on the
upper-right have been obtained keeping only the events that fulfil the
veto condition. It can be seen that the mismatch between DR and DS in
the high-$\pt$ tail is less sizeable when the $b$-veto is applied,
which is the expected result. Since in ref.~\cite{Frixione:2008yi} it
was noticed that the transverse momentum of the system made by the two
hardest leptons turns out to be sensitive to the treatment of the
doubly-resonant region, in the lower panel of
fig.~\ref{fig:cmpbveto_POWHER-DR-DS} we show also the predictions for
$\pt^{(\ell_1 \ell_2\!)}$.  Also for this quantity the effect of the
$b$-veto is to reduce the differences between DR and DS, as the plot
in the lower-right panel shows, in accordance with what has been found
in ref.~\cite{Frixione:2008yi}.

The conclusions of this analysis can be summarized as follows. We
checked that the $b$-veto reduces the differences between DR and
DS. This is the expected result, since the $b$-veto was originally
proposed by the authors of ref.~\cite{Campbell:2005bb} to reduce
interference effects from $t\bar{t}$ production. In fact, by requiring
a $b$-veto, one reduces the number of events with two hard and central
$b$-flavoured hadrons, which is indeed one of the typical signatures
of $t\bar{t}$ production.  The fact that DR and DS total cross
sections become closer when the $b$-veto is applied is also in
accordance with the interpretation of the difference between DR and DS
being a measure of the interference between $Wt$ and $t\bar{t}$
production. This interpretation is supported by the plots in
fig.~\ref{fig:cmpbveto_POWHER-DR-DS}, where it is shown that the
difference between DR and DS is reduced also for differential
distributions, when the $b$-veto is applied. However, the plots in
fig.~\ref{fig:cmpbveto_POWHER-DR-DS} show also that observables are
potentially affected by the $b$-veto non uniformly (i.e.~ratios
between DR and DS results can be bin-dependent). This suggests that
particular care should be taken when one performs a full analysis
where the contribution from the $Wt$-channel process is supposed to be
relevant, since the size of interference effects may depend on the
cuts applied. As it was already observed by the \MCatNLO{} authors, a
comparison between DR and DS predictions gives an estimate of the
theoretical uncertainty due to these effects.

\begin{figure}[!htb]
  \begin{center}
    \epsfig{file=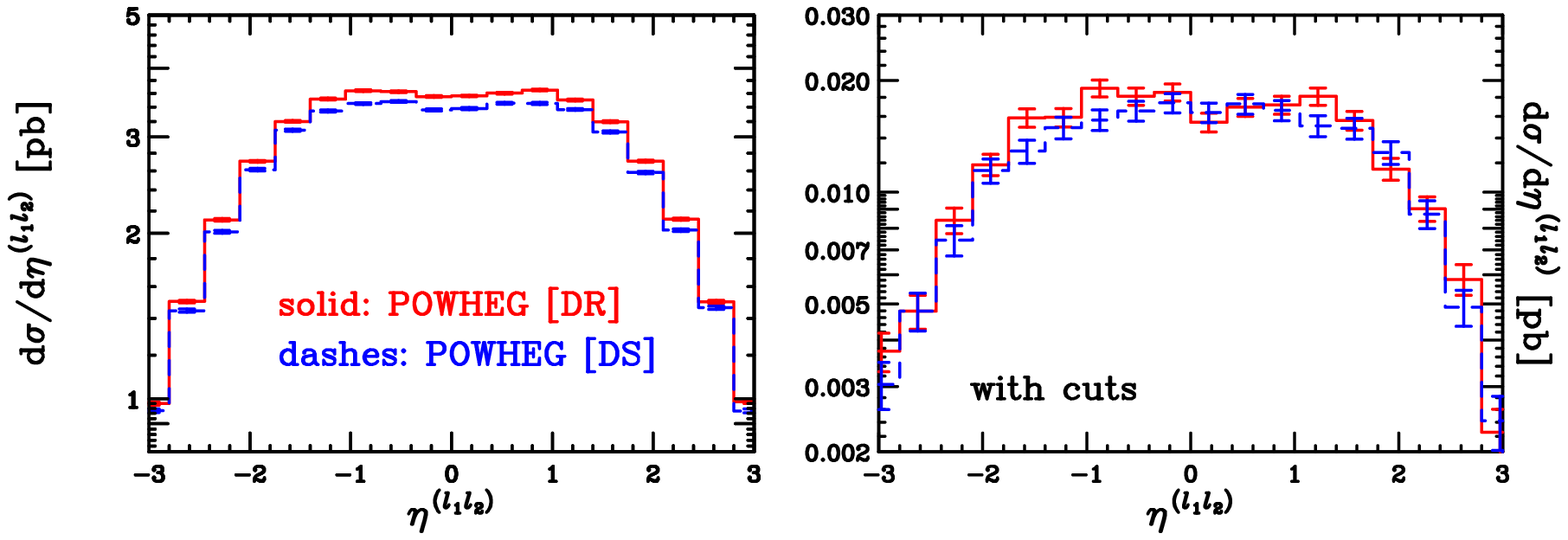,width=\figwidth}\\
    \epsfig{file=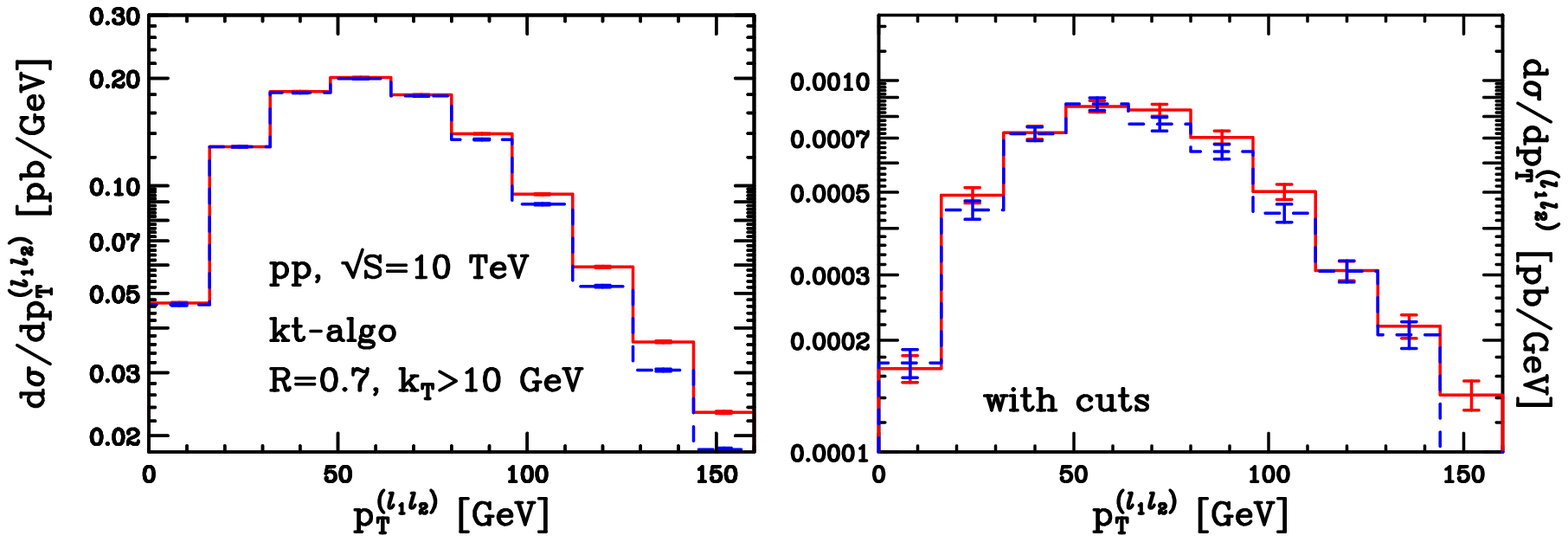,width=\figwidth}
  \end{center}
  \captskip
  \caption{\label{fig:cmp1cuts1_POWPYT-DR-DS} Comparisons between
    \POWHEG{} (interfaced to \PYTHIA{}) results obtained with the DR
    and DS prescriptions, for leptonic quantities.  Plots in the right
    panel have been obtained with the ``$Wt$-cuts'' described in the
    text. Uncertainties indicated by the vertical bars are only
    statistical. }
\end{figure}

\begin{figure}[htb]
  \begin{center}
    \epsfig{file=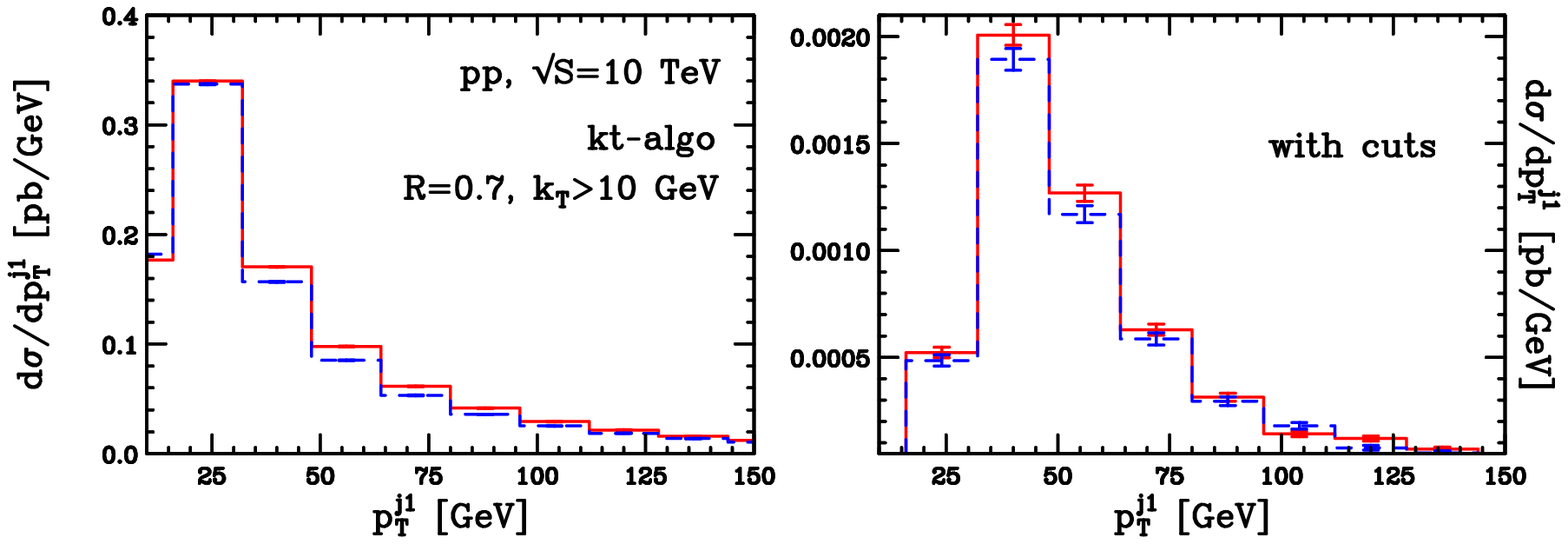,width=\figwidth}\\
    \epsfig{file=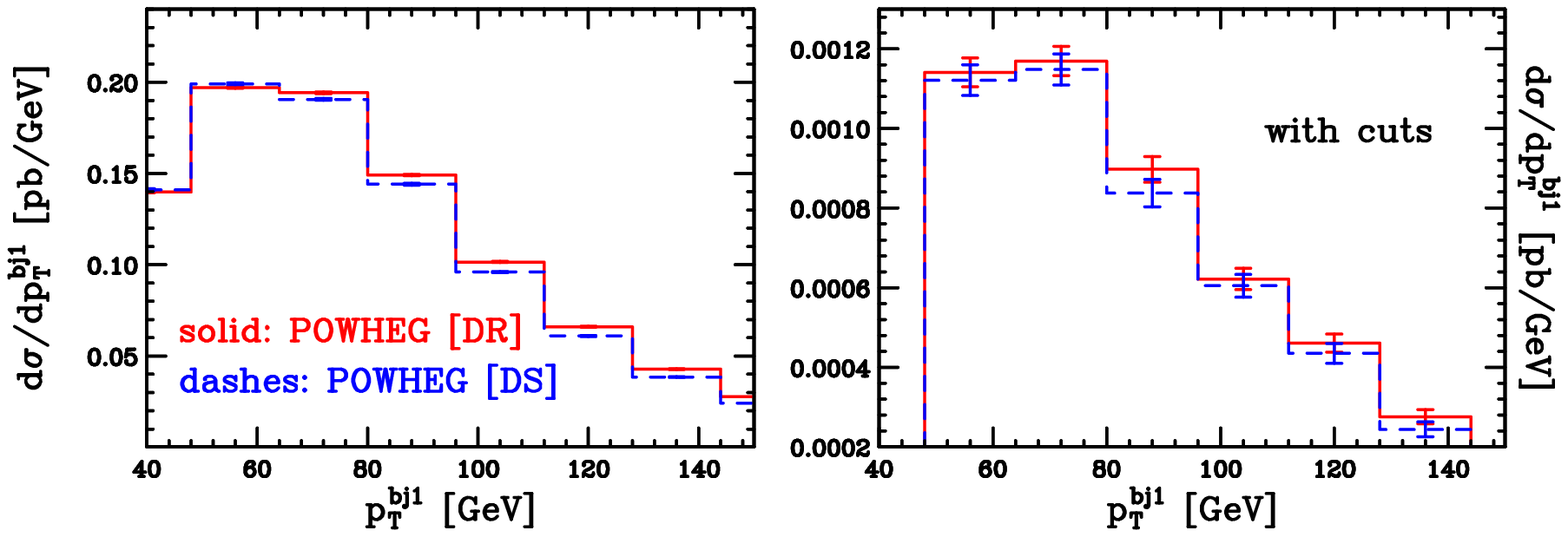,width=\figwidth}
  \end{center}
  \captskip
  \caption{\label{fig:cmp2cuts1_POWPYT-DR-DS} Comparisons between
    \POWHEG{} (interfaced to \PYTHIA{}) results obtained with the DR
    and DS prescriptions.  Plots in the right panel have been obtained
    with the ``$Wt$-cuts'' described in the text. Uncertainties
    indicated by the vertical bars are only statistical.}
\end{figure}

We have also performed some comparisons between DR and DS results
obtained with the \PYTHIA{} shower.  In order to maximize consistency
with the \POWHEG{} prescriptions, we used the $\pt$-ordered shower. In
fig.~\ref{fig:cmp1cuts1_POWPYT-DR-DS} we show the pseudorapidity and
the transverse momentum of the system made by the two hardest leptons,
while in fig.~\ref{fig:cmp2cuts1_POWPYT-DR-DS} the transverse momentum
of the hardest non $b$-flavoured jet ($\pt^{\,j_1}$) and for the
hardest jet that contains a $b$-flavoured hadron ($\pt^{\,bj_1}$) are
shown.  The plots on the left have been obtained without imposing any
cut. Instead, the plots on the right have been obtained using cuts
similar to the ``$Wt$-signal cuts'' of ref.~\cite{White:2009yt}. Jets
have been defined according to the $\kt$
algorithm~\cite{Catani:1993hr}, as implemented in the {\tt FASTJET}
package~\cite{Cacciari:2005hq}, setting $R=0.7$ and imposing a lower
$10$~GeV cut on jet transverse momenta.  To accept an event, we
required the following properties:
\begin{itemize}
\item There is exactly one $b$-jet with $\pt^j>50$ GeV and
  $|\eta^j|<2.5$. A $b$-jet is defined as a jet that contains at least
  one $b$-flavoured hadron and has $\pt^j>25$ GeV and $|\eta^j|<2.5$.
\item There are at least two light-flavoured jets with $\pt^j>25$ GeV
  and $|\eta^j|<2.5$. The invariant mass of the system made by the two
  hardest jets among these light-flavoured jets has to lie within $55$
  and $85$ GeV.
\item There is one lepton with $\pt^{\ell}>25$ GeV and
  $|\eta^{\ell}|<2.5$.  This lepton has also to be isolated with
  respect to the $b$-jet and the two light-flavoured jets, i.e.~its
  distance from the jets in the $(\eta,\phi)$ plane has to be larger
  than $0.4$.
\item The missing transverse energy is larger than $25$ GeV.
\end{itemize}
Although we have not performed as detailed a study as the one of
ref.~\cite{White:2009yt}, from figs.~\ref{fig:cmp1cuts1_POWPYT-DR-DS}
and~\ref{fig:cmp2cuts1_POWPYT-DR-DS} we observe that the DR and the DS
predictions are consistent (within the statistical accuracy) also when
the above cuts are applied, as was observed in the aforementioned
work.

\cleardoublepage
\section{Conclusions}
\label{sec:conclusions}
In this paper we have described the implementation of $Wt$-channel
single-top production at next-to-leading order in QCD, in the
\POWHEG{} framework. We have used the \BOX{} package, which is a
program that automates the algorithm first proposed in
ref.~\cite{Nason:2004rx} and then carefully described in
ref.~\cite{Frixione:2007vw}.  Having used this package, we described
how we calculated the main needed ingredients.

Since NLO corrections to single-top production in the $Wt$-channel are
known not to be well-defined (real contributions interfere with
$t\bar{t}$ production), we decided to follow the same strategy
originally proposed by the \MCatNLO{} authors in
ref.~\cite{Frixione:2008yi}: we included two definitions of the NLO
corrections, known as DR (\emph{Diagram Removal}) and DS
(\emph{Diagram Subtraction}), both of which can be used when the
merging of the fixed order result with parton showers is performed.
Moreover, the difference of results obtained with these two
prescriptions gives an estimate of the size of interference
effects. We have described how we included the two prescriptions in
\POWHEG{}, and how we dealt with DS and the problem of its exact
implementation within the \POWHEG{} method.

To check the correctness of the whole implementation and to assess the
validity of the choices we made, results have been compared with the
\MCatNLO{} program for the LHC, where $Wt$ production is relevant: we
found very good agreement, both for DR and DS.
%
%
We have also compared DR and DS results when a $b$-veto is imposed. We
found that a $b$-veto reduces interference effects with $t\bar{t}$,
the difference between DR and DS results becoming smaller when the
veto is in place.  Moreover, we have also presented some results
obtained with the \PYTHIA{} shower and with typical defining cuts for
the $Wt$ signal. Although we have not performed as full an analysis as
the one reported in ref.~\cite{White:2009yt}, good agreement between
DR and DS has been found also when typical $Wt$-channel cuts are
applied.

The main purpose of this work was the completion of the work presented
in ref.~\cite{Alioli:2009je}, where the \POWHEG{} implementation of
single-top $s$- and $t$-channel was described. Therefore, at present
all single-top processes can be simulated in the context of a NLO+PS
approach with \POWHEG{} as well as with \MCatNLO{}. We also stress
that, in \POWHEG{}, $Wt$-channel was the missing process among the
ones relevant for $H\to WW$ searches: Higgs production via
gluon~\cite{Alioli:2008tz} and vector-boson fusion~\cite{Nason:2009ai}
are already available, as well as $t\bar{t}$~\cite{Frixione:2007nw}
and $VV$~\cite{Hamilton:VV} production, which (together with
single-top $Wt$) are the main backgrounds .

The computer code for this \POWHEG{} implementation will soon be
available within the public branch of the \BOX{} package, that
can be downloaded at the site
\begin{center}
  \url{http://virgilio.mib.infn.it/~nason/POWHEG}
\end{center}

\section*{Acknowledgements}
I would like to particularly thank C. White, for providing me with the
one-loop corrections for this process, for many useful discussions on
this project, and also for comments on the manuscript.  I would also
like to thank R. Frederix and G. Zanderighi for helpful hints on the
usage of MadGraph and QCDloop.  I am also grateful to S. Alioli and
K. Hamilton for comments on the manuscript.

\bibliography{paper}

\providecommand{\href}[2]{#2}\begingroup\raggedright\begin{thebibliography}{10}

\bibitem{Aaltonen:2009jj}
{\bf CDF} Collaboration, T.~Aaltonen {\em et~al.}, {\it {First Observation of
  Electroweak Single Top Quark Production}},  {\em Phys. Rev. Lett.} {\bf 103}
  (2009) 092002, [\href{http://xxx.lanl.gov/abs/0903.0885}{{\tt 0903.0885}}].

\bibitem{Abazov:2009ii}
{\bf D0} Collaboration, V.~M. Abazov {\em et~al.}, {\it {Observation of Single
  Top-Quark Production}},  {\em Phys. Rev. Lett.} {\bf 103} (2009) 092001,
  [\href{http://xxx.lanl.gov/abs/0903.0850}{{\tt 0903.0850}}].

\bibitem{Alwall:2006bx}
J.~Alwall {\em et~al.}, {\it {Is $V_{tb} = 1$?}},  {\em Eur. Phys. J.} {\bf
  C49} (2007) 791--801, [\href{http://xxx.lanl.gov/abs/hep-ph/0607115}{{\tt
  hep-ph/0607115}}].

\bibitem{Mahlon:1996pn}
G.~Mahlon and S.~J. Parke, {\it {Improved spin basis for angular correlation
  studies in single top quark production at the Tevatron}},  {\em Phys. Rev.}
  {\bf D55} (1997) 7249--7254,
  [\href{http://xxx.lanl.gov/abs/hep-ph/9611367}{{\tt hep-ph/9611367}}].

\bibitem{Mahlon:1999gz}
G.~Mahlon and S.~J. Parke, {\it {Single top quark production at the LHC:
  Understanding spin}},  {\em Phys. Lett.} {\bf B476} (2000) 323--330,
  [\href{http://xxx.lanl.gov/abs/hep-ph/9912458}{{\tt hep-ph/9912458}}].

\bibitem{Tait:2000sh}
T.~M.~P. Tait and C.~P. Yuan, {\it {Single top quark production as a window to
  physics beyond the standard model}},  {\em Phys. Rev.} {\bf D63} (2001)
  014018, [\href{http://xxx.lanl.gov/abs/hep-ph/0007298}{{\tt
  hep-ph/0007298}}].

\bibitem{Cao:2007ea}
Q.-H. Cao, J.~Wudka, and C.~P. Yuan, {\it {Search for New Physics via Single
  Top Production at the LHC}},  {\em Phys. Lett.} {\bf B658} (2007) 50--56,
  [\href{http://xxx.lanl.gov/abs/0704.2809}{{\tt 0704.2809}}].

\bibitem{Dittmar:1996ss}
M.~Dittmar and H.~K. Dreiner, {\it {How to find a Higgs boson with a mass
  between 155-GeV - 180-GeV at the LHC}},  {\em Phys. Rev.} {\bf D55} (1997)
  167--172, [\href{http://xxx.lanl.gov/abs/hep-ph/9608317}{{\tt
  hep-ph/9608317}}].

\bibitem{Bordes:1994ki}
G.~Bordes and B.~van Eijk, {\it {Calculating QCD corrections to single top
  production in hadronic interactions}},  {\em Nucl. Phys.} {\bf B435} (1995)
  23--58.

\bibitem{Giele:1995kr}
W.~T. Giele, S.~Keller, and E.~Laenen, {\it {QCD corrections to $W$ boson plus
  heavy quark production at the Tevatron}},  {\em Phys. Lett.} {\bf B372}
  (1996) 141--149, [\href{http://xxx.lanl.gov/abs/hep-ph/9511449}{{\tt
  hep-ph/9511449}}].

\bibitem{Stelzer:1997ns}
T.~Stelzer, Z.~Sullivan, and S.~Willenbrock, {\it {Single top quark production
  via $W$-gluon fusion at next-to-leading order}},  {\em Phys. Rev.} {\bf D56}
  (1997) 5919--5927, [\href{http://xxx.lanl.gov/abs/hep-ph/9705398}{{\tt
  hep-ph/9705398}}].

\bibitem{Harris:2002md}
B.~W. Harris, E.~Laenen, L.~Phaf, Z.~Sullivan, and S.~Weinzierl, {\it {The
  Fully differential single top quark cross-section in next to leading order
  QCD}},  {\em Phys. Rev.} {\bf D66} (2002) 054024,
  [\href{http://xxx.lanl.gov/abs/hep-ph/0207055}{{\tt hep-ph/0207055}}].

\bibitem{Zhu:2002uj}
S.~Zhu, {\it {Next-to-leading order QCD corrections to b g $\to$ t W- at the
  CERN Large Hadron Collider}},  {\em Phys. Lett.} {\bf B524} (2002) 283--288.

\bibitem{Campbell:2004ch}
J.~M. Campbell, R.~K. Ellis, and F.~Tramontano, {\it {Single top production and
  decay at next-to-leading order}},  {\em Phys. Rev.} {\bf D70} (2004) 094012,
  [\href{http://xxx.lanl.gov/abs/hep-ph/0408158}{{\tt hep-ph/0408158}}].

\bibitem{Campbell:2005bb}
J.~M. Campbell and F.~Tramontano, {\it {Next-to-leading order corrections to W
  t production and decay}},  {\em Nucl. Phys.} {\bf B726} (2005) 109--130,
  [\href{http://xxx.lanl.gov/abs/hep-ph/0506289}{{\tt hep-ph/0506289}}].

\bibitem{Cao:2004ap}
Q.-H. Cao, R.~Schwienhorst, and C.~P. Yuan, {\it {Next-to-leading order
  corrections to single top quark production and decay at Tevatron. 1.
  $s$-channel process}},  {\em Phys. Rev.} {\bf D71} (2005) 054023,
  [\href{http://xxx.lanl.gov/abs/hep-ph/0409040}{{\tt hep-ph/0409040}}].

\bibitem{Cao:2005pq}
Q.-H. Cao, R.~Schwienhorst, J.~A. Benitez, R.~Brock, and C.~P. Yuan, {\it
  {Next-to-leading order corrections to single top quark production and decay
  at the Tevatron: 2. $t$-channel process}},  {\em Phys. Rev.} {\bf D72} (2005)
  094027, [\href{http://xxx.lanl.gov/abs/hep-ph/0504230}{{\tt
  hep-ph/0504230}}].

\bibitem{Kidonakis:2006bu}
N.~Kidonakis, {\it {Single top production at the Tevatron: Threshold
  resummation and finite-order soft gluon corrections}},  {\em Phys. Rev.} {\bf
  D74} (2006) 114012, [\href{http://xxx.lanl.gov/abs/hep-ph/0609287}{{\tt
  hep-ph/0609287}}].

\bibitem{Campbell:2009ss}
J.~M. Campbell, R.~Frederix, F.~Maltoni, and F.~Tramontano, {\it
  {Next-to-Leading-Order Predictions for t-Channel Single-Top Production at
  Hadron Colliders}},  {\em Phys. Rev. Lett.} {\bf 102} (2009) 182003,
  [\href{http://xxx.lanl.gov/abs/0903.0005}{{\tt 0903.0005}}].

\bibitem{Campbell:2009gj}
J.~M. Campbell, R.~Frederix, F.~Maltoni, and F.~Tramontano, {\it {NLO
  predictions for t-channel production of single top and fourth generation
  quarks at hadron colliders}},  {\em JHEP} {\bf 10} (2009) 042,
  [\href{http://xxx.lanl.gov/abs/0907.3933}{{\tt 0907.3933}}].

\bibitem{Frixione:2002ik}
S.~Frixione and B.~R. Webber, {\it {Matching NLO QCD computations and parton
  shower simulations}},  {\em JHEP} {\bf 06} (2002) 029,
  [\href{http://xxx.lanl.gov/abs/hep-ph/0204244}{{\tt hep-ph/0204244}}].

\bibitem{Nason:2004rx}
P.~Nason, {\it {A new method for combining NLO QCD with shower Monte Carlo
  algorithms}},  {\em JHEP} {\bf 11} (2004) 040,
  [\href{http://xxx.lanl.gov/abs/hep-ph/0409146}{{\tt hep-ph/0409146}}].

\bibitem{Frixione:2007vw}
S.~Frixione, P.~Nason, and C.~Oleari, {\it {Matching NLO QCD computations with
  Parton Shower simulations: the POWHEG method}},  {\em JHEP} {\bf 11} (2007)
  070, [\href{http://xxx.lanl.gov/abs/0709.2092}{{\tt 0709.2092}}].

\bibitem{Nagy:2005aa}
Z.~Nagy and D.~E. Soper, {\it Matching parton showers to nlo computations},
  {\em JHEP} {\bf 10} (2005) 024,
  [\href{http://xxx.lanl.gov/abs/hep-ph/0503053}{{\tt hep-ph/0503053}}].

\bibitem{Giele:2007di}
W.~T. Giele, D.~A. Kosower, and P.~Z. Skands, {\it {A Simple shower and
  matching algorithm}},  {\em Phys. Rev.} {\bf D78} (2008) 014026,
  [\href{http://xxx.lanl.gov/abs/0707.3652}{{\tt 0707.3652}}].

\bibitem{Lavesson:2008ah}
N.~Lavesson and L.~Lonnblad, {\it {Extending CKKW-merging to One-Loop Matrix
  Elements}},  {\em JHEP} {\bf 12} (2008) 070,
  [\href{http://xxx.lanl.gov/abs/0811.2912}{{\tt 0811.2912}}].

\bibitem{LatundeDada:2006gx}
O.~Latunde-Dada, S.~Gieseke, and B.~Webber, {\it {A positive-weight
  next-to-leading-order Monte Carlo for $e^+ e^-$ annihilation to hadrons}},
  {\em JHEP} {\bf 02} (2007) 051,
  [\href{http://xxx.lanl.gov/abs/hep-ph/0612281}{{\tt hep-ph/0612281}}].

\bibitem{LatundeDada:2007jg}
O.~Latunde-Dada, {\it {Herwig Monte Carlo At Next-To-Leading Order for $e^+
  e^-$ annihilation and lepton pair production}},  {\em JHEP} {\bf 11} (2007)
  040, [\href{http://xxx.lanl.gov/abs/0708.4390}{{\tt 0708.4390}}].

\bibitem{LatundeDada:2008bv}
O.~Latunde-Dada, {\it {Applying the POWHEG method to top pair production and
  decays at the ILC}},  {\em Eur. Phys. J.} {\bf C58} (2008) 543--554,
  [\href{http://xxx.lanl.gov/abs/0806.4560}{{\tt 0806.4560}}].

\bibitem{Frixione:2003ei}
S.~Frixione, P.~Nason, and B.~R. Webber, {\it {Matching NLO QCD and parton
  showers in heavy flavour production}},  {\em JHEP} {\bf 08} (2003) 007,
  [\href{http://xxx.lanl.gov/abs/hep-ph/0305252}{{\tt hep-ph/0305252}}].

\bibitem{Frixione:2005vw}
S.~Frixione, E.~Laenen, P.~Motylinski, and B.~R. Webber, {\it {Single-top
  production in MC@NLO}},  {\em JHEP} {\bf 03} (2006) 092,
  [\href{http://xxx.lanl.gov/abs/hep-ph/0512250}{{\tt hep-ph/0512250}}].

\bibitem{Nason:2006hf}
P.~Nason and G.~Ridolfi, {\it {A positive-weight next-to-leading-order Monte
  Carlo for $Z$ pair hadroproduction}},  {\em JHEP} {\bf 08} (2006) 077,
  [\href{http://xxx.lanl.gov/abs/hep-ph/0606275}{{\tt hep-ph/0606275}}].

\bibitem{Frixione:2007nw}
S.~Frixione, P.~Nason, and G.~Ridolfi, {\it {A Positive-Weight
  Next-to-Leading-Order Monte Carlo for Heavy Flavour Hadroproduction}},  {\em
  JHEP} {\bf 09} (2007) 126, [\href{http://xxx.lanl.gov/abs/0707.3088}{{\tt
  0707.3088}}].

\bibitem{Alioli:2008gx}
S.~Alioli, P.~Nason, C.~Oleari, and E.~Re, {\it {NLO vector-boson production
  matched with shower in POWHEG}},  {\em JHEP} {\bf 07} (2008) 060,
  [\href{http://xxx.lanl.gov/abs/0805.4802}{{\tt 0805.4802}}].

\bibitem{Hamilton:2008pd}
K.~Hamilton, P.~Richardson, and J.~Tully, {\it {A Positive-Weight
  Next-to-Leading Order Monte Carlo Simulation of Drell-Yan Vector Boson
  Production}},  {\em JHEP} {\bf 10} (2008) 015,
  [\href{http://xxx.lanl.gov/abs/0806.0290}{{\tt 0806.0290}}].

\bibitem{Frixione:2008yi}
S.~Frixione, E.~Laenen, P.~Motylinski, B.~R. Webber, and C.~D. White, {\it
  {Single-top hadroproduction in association with a W boson}},  {\em JHEP} {\bf
  07} (2008) 029, [\href{http://xxx.lanl.gov/abs/0805.3067}{{\tt 0805.3067}}].

\bibitem{Alioli:2008tz}
S.~Alioli, P.~Nason, C.~Oleari, and E.~Re, {\it {NLO Higgs boson production via
  gluon fusion matched with shower in POWHEG}},  {\em JHEP} {\bf 04} (2009)
  002, [\href{http://xxx.lanl.gov/abs/0812.0578}{{\tt 0812.0578}}].

\bibitem{Hamilton:2009za}
K.~Hamilton, P.~Richardson, and J.~Tully, {\it {A Positive-Weight
  Next-to-Leading Order Monte Carlo Simulation for Higgs Boson Production}},
  {\em JHEP} {\bf 04} (2009) 116,
  [\href{http://xxx.lanl.gov/abs/0903.4345}{{\tt 0903.4345}}].

\bibitem{Papaefstathiou:2009sr}
A.~Papaefstathiou and O.~Latunde-Dada, {\it {NLO production of $W$ ' bosons at
  hadron colliders using the MC@NLO and POWHEG methods}},  {\em JHEP} {\bf 07}
  (2009) 044, [\href{http://xxx.lanl.gov/abs/0901.3685}{{\tt 0901.3685}}].

\bibitem{LatundeDada:2009rr}
O.~Latunde-Dada, {\it {MC@NLO for the hadronic decay of Higgs bosons in
  associated production with vector bosons}},  {\em JHEP} {\bf 05} (2009) 112,
  [\href{http://xxx.lanl.gov/abs/0903.4135}{{\tt 0903.4135}}].

\bibitem{Alioli:2009je}
S.~Alioli, P.~Nason, C.~Oleari, and E.~Re, {\it {NLO single-top production
  matched with shower in POWHEG: s- and t-channel contributions}},  {\em JHEP}
  {\bf 09} (2009) 111, [\href{http://xxx.lanl.gov/abs/0907.4076}{{\tt
  0907.4076}}].

\bibitem{Nason:2009ai}
P.~Nason and C.~Oleari, {\it {NLO Higgs boson production via vector-boson
  fusion matched with shower in POWHEG}},  {\em JHEP} {\bf 02} (2010) 037,
  [\href{http://xxx.lanl.gov/abs/0911.5299}{{\tt 0911.5299}}].

\bibitem{Weydert:2009vr}
C.~Weydert {\em et~al.}, {\it {Charged Higgs boson production in association
  with a top quark in MC@NLO}},  {\em Eur. Phys. J.} {\bf C67} (2010) 617--636,
  [\href{http://xxx.lanl.gov/abs/0912.3430}{{\tt 0912.3430}}].

\bibitem{Torrielli:2010aw}
P.~Torrielli and S.~Frixione, {\it {Matching NLO QCD computations with PYTHIA
  using MC@NLO}},  {\em JHEP} {\bf 04} (2010) 110,
  [\href{http://xxx.lanl.gov/abs/1002.4293}{{\tt 1002.4293}}].

\bibitem{Corcella:2000bw}
G.~Corcella {\em et~al.}, {\it {HERWIG 6.5: an event generator for Hadron
  Emission Reactions With Interfering Gluons (including supersymmetric
  processes)}},  {\em JHEP} {\bf 01} (2001) 010,
  [\href{http://xxx.lanl.gov/abs/hep-ph/0011363}{{\tt hep-ph/0011363}}].

\bibitem{Sjostrand:2006za}
T.~Sjostrand, S.~Mrenna, and P.~Skands, {\it Pythia 6.4 physics and manual},
  {\em JHEP} {\bf 05} (2006) 026,
  [\href{http://xxx.lanl.gov/abs/hep-ph/0603175}{{\tt hep-ph/0603175}}].

\bibitem{Bahr:2008pv}
M.~Bahr {\em et~al.}, {\it {Herwig++ Physics and Manual}},  {\em Eur. Phys. J.}
  {\bf C58} (2008) 639--707, [\href{http://xxx.lanl.gov/abs/0803.0883}{{\tt
  0803.0883}}].

\bibitem{Butterworth:2010ym}
J.~M. Butterworth {\em et~al.}, {\it {The Tools and Monte Carlo working group
  Summary Report}},  \href{http://xxx.lanl.gov/abs/1003.1643}{{\tt 1003.1643}}.

\bibitem{Hoeche:2010pf}
S.~Hoeche, F.~Krauss, M.~Schonherr, and F.~Siegert, {\it {Automating the POWHEG
  method in Sherpa}},  \href{http://xxx.lanl.gov/abs/1008.5399}{{\tt
  1008.5399}}.

\bibitem{Nason:2010ap}
P.~Nason, {\it {Recent developments in POWHEG}},  {\em PoS} {\bf RADCOR2009}
  (2010) 018, [\href{http://xxx.lanl.gov/abs/1001.2747}{{\tt 1001.2747}}].

\bibitem{Alioli:2010xd}
S.~Alioli, P.~Nason, C.~Oleari, and E.~Re, {\it {A general framework for
  implementing NLO calculations in shower Monte Carlo programs: the POWHEG
  BOX}},  {\em JHEP} {\bf 06} (2010) 043,
  [\href{http://xxx.lanl.gov/abs/1002.2581}{{\tt 1002.2581}}].

\bibitem{Belyaev:1998dn}
A.~S. Belyaev, E.~E. Boos, and L.~V. Dudko, {\it {Single top quark at future
  hadron colliders: Complete signal and background study}},  {\em Phys. Rev.}
  {\bf D59} (1999) 075001, [\href{http://xxx.lanl.gov/abs/hep-ph/9806332}{{\tt
  hep-ph/9806332}}].

\bibitem{Tait:1999cf}
T.~M.~P. Tait, {\it {The $t W^{-}$ mode of single top production}},  {\em Phys.
  Rev.} {\bf D61} (2000) 034001,
  [\href{http://xxx.lanl.gov/abs/hep-ph/9909352}{{\tt hep-ph/9909352}}].

\bibitem{White:2009yt}
C.~D. White, S.~Frixione, E.~Laenen, and F.~Maltoni, {\it {Isolating Wt
  production at the LHC}},  {\em JHEP} {\bf 11} (2009) 074,
  [\href{http://xxx.lanl.gov/abs/0908.0631}{{\tt 0908.0631}}].

\bibitem{Alwall:2007st}
J.~Alwall {\em et~al.}, {\it {MadGraph/MadEvent v4: The New Web Generation}},
  {\em JHEP} {\bf 09} (2007) 028,
  [\href{http://xxx.lanl.gov/abs/0706.2334}{{\tt 0706.2334}}].

\bibitem{Mertig:1990an}
R.~Mertig, M.~Bohm, and A.~Denner, {\it {FEYN CALC: Computer algebraic
  calculation of Feynman amplitudes}},  {\em Comput. Phys. Commun.} {\bf 64}
  (1991) 345--359.

\bibitem{Frixione:1995ms}
S.~Frixione, Z.~Kunszt, and A.~Signer, {\it {Three-jet cross sections to
  next-to-leading order}},  {\em Nucl. Phys.} {\bf B467} (1996) 399--442,
  [\href{http://xxx.lanl.gov/abs/hep-ph/9512328}{{\tt hep-ph/9512328}}].

\bibitem{Frixione:1997np}
S.~Frixione, {\it {A general approach to jet cross sections in QCD}},  {\em
  Nucl. Phys.} {\bf B507} (1997) 295--314,
  [\href{http://xxx.lanl.gov/abs/hep-ph/9706545}{{\tt hep-ph/9706545}}].

\bibitem{Ellis:2007qk}
R.~K. Ellis and G.~Zanderighi, {\it {Scalar one-loop integrals for QCD}},  {\em
  JHEP} {\bf 02} (2008) 002, [\href{http://xxx.lanl.gov/abs/0712.1851}{{\tt
  0712.1851}}].

\bibitem{Pittau:1996rp}
R.~Pittau, {\it {Final state QCD corrections to off-shell single top production
  in hadron collisions}},  {\em Phys. Lett.} {\bf B386} (1996) 397--402,
  [\href{http://xxx.lanl.gov/abs/hep-ph/9603265}{{\tt hep-ph/9603265}}].

\bibitem{Falgari:2010sf}
P.~Falgari, P.~Mellor, and A.~Signer, {\it {Production-decay interferences at
  NLO in QCD for t-channel single-top production}},
  \href{http://xxx.lanl.gov/abs/1007.0893}{{\tt 1007.0893}}.

\bibitem{Pumplin:2002vw}
J.~Pumplin {\em et~al.}, {\it {New generation of parton distributions with
  uncertainties from global QCD analysis}},  {\em JHEP} {\bf 07} (2002) 012,
  [\href{http://xxx.lanl.gov/abs/hep-ph/0201195}{{\tt hep-ph/0201195}}].

\bibitem{Frixione:2007zp}
S.~Frixione, E.~Laenen, P.~Motylinski, and B.~R. Webber, {\it {Angular
  correlations of lepton pairs from vector boson and top quark decays in Monte
  Carlo simulations}},  {\em JHEP} {\bf 04} (2007) 081,
  [\href{http://xxx.lanl.gov/abs/hep-ph/0702198}{{\tt hep-ph/0702198}}].

\bibitem{Catani:1993hr}
S.~Catani, Y.~L. Dokshitzer, M.~H. Seymour, and B.~R. Webber, {\it
  {Longitudinally invariant $K_t$ clustering algorithms for hadron hadron
  collisions}},  {\em Nucl. Phys.} {\bf B406} (1993) 187--224.

\bibitem{Cacciari:2005hq}
M.~Cacciari and G.~P. Salam, {\it {Dispelling the $N^{3}$ myth for the $k_t$
  jet-finder}},  {\em Phys. Lett.} {\bf B641} (2006) 57--61,
  [\href{http://xxx.lanl.gov/abs/hep-ph/0512210}{{\tt hep-ph/0512210}}].

\bibitem{Hamilton:VV}
K.~Hamilton.~{\rm To appear}.

\end{thebibliography}\endgroup

\end{document}